\documentclass[preprint,authoryear,12pt]{elsarticle}

\usepackage[utf8]{inputenc}
\usepackage{amsmath}
\usepackage{graphicx}
\usepackage{booktabs}
\usepackage{multirow}
\usepackage{caption}
\usepackage{subcaption}

\usepackage{hyperref}

\begin{document}

% ======================================================================
% IMPORTANT: Include auto-generated macros
% ======================================================================
% Auto-generated LaTeX macros from experimental results
% Generated: 2025-12-20 17:25:38
% Source: parse_results.py
% DO NOT EDIT MANUALLY - Run 'make macros' to regenerate
%
% Total macros: 427

% ADTV Statistics (Table 1-2)
\newcommand{\adtvObservations}{2,753,106}

% Full Sample Event Study (Table 4)
\newcommand{\allSamplesDayMinusFiveAAR}{-0.07\%}
\newcommand{\allSamplesDayMinusFiveCAR}{-4.93\%}
\newcommand{\allSamplesDayMinusFiveTstat}{-5.12}
\newcommand{\allSamplesDayMinusOneAAR}{0.66\%}
\newcommand{\allSamplesDayMinusOneCAR}{-4.09\%}
\newcommand{\allSamplesDayMinusOneTstat}{28.87}
\newcommand{\allSamplesDayMinusTenAAR}{-0.10\%}
\newcommand{\allSamplesDayMinusTenCAR}{-4.46\%}
\newcommand{\allSamplesDayMinusTenTstat}{-6.93}
\newcommand{\allSamplesDayMinusTwentyFiveAAR}{-0.13\%}
\newcommand{\allSamplesDayMinusTwentyFiveCAR}{-2.09\%}
\newcommand{\allSamplesDayMinusTwentyFiveTstat}{-7.52}
\newcommand{\allSamplesDayMinusTwoAAR}{0.14\%}
\newcommand{\allSamplesDayMinusTwoCAR}{-4.75\%}
\newcommand{\allSamplesDayMinusTwoTstat}{8.28}
\newcommand{\allSamplesDayPlusFiftyAAR}{0.02\%}
\newcommand{\allSamplesDayPlusFiftyCAR}{3.09\%}
\newcommand{\allSamplesDayPlusFiftyTstat}{1.01}
\newcommand{\allSamplesDayPlusFiveAAR}{-0.05\%}
\newcommand{\allSamplesDayPlusFiveCAR}{2.19\%}
\newcommand{\allSamplesDayPlusFiveTstat}{-2.34}
\newcommand{\allSamplesDayPlusOneAAR}{0.09\%}
\newcommand{\allSamplesDayPlusOneCAR}{2.15\%}
\newcommand{\allSamplesDayPlusOneTstat}{2.72}
\newcommand{\allSamplesDayPlusTenAAR}{0.01\%}
\newcommand{\allSamplesDayPlusTenCAR}{2.33\%}
\newcommand{\allSamplesDayPlusTenTstat}{0.28}
\newcommand{\allSamplesDayPlusTwentyFiveAAR}{0.03\%}
\newcommand{\allSamplesDayPlusTwentyFiveCAR}{2.71\%}
\newcommand{\allSamplesDayPlusTwentyFiveTstat}{1.28}
\newcommand{\allSamplesDayPlusTwoAAR}{-0.00\%}
\newcommand{\allSamplesDayPlusTwoCAR}{2.15\%}
\newcommand{\allSamplesDayPlusTwoTstat}{-0.18}
\newcommand{\allSamplesDayZeroAAR}{6.15\%}
\newcommand{\allSamplesDayZeroCAR}{2.06\%}
\newcommand{\allSamplesDayZeroTstat}{118.81}

% Firm Size Analysis (Table 5)
\newcommand{\largecapDayMinusFiveAAR}{0.06\%}
\newcommand{\largecapDayMinusFiveCAR}{-3.37\%}
\newcommand{\largecapDayMinusFiveTstat}{0.61}
\newcommand{\largecapDayMinusOneAAR}{0.73\%}
\newcommand{\largecapDayMinusOneCAR}{-1.94\%}
\newcommand{\largecapDayMinusOneTstat}{4.86}
\newcommand{\largecapDayMinusTenAAR}{-0.10\%}
\newcommand{\largecapDayMinusTenCAR}{-3.51\%}
\newcommand{\largecapDayMinusTenTstat}{-1.04}
\newcommand{\largecapDayMinusTwentyFiveAAR}{-0.04\%}
\newcommand{\largecapDayMinusTwentyFiveCAR}{-2.29\%}
\newcommand{\largecapDayMinusTwentyFiveTstat}{-0.41}
\newcommand{\largecapDayMinusTwoAAR}{0.38\%}
\newcommand{\largecapDayMinusTwoCAR}{-2.67\%}
\newcommand{\largecapDayMinusTwoTstat}{3.29}
\newcommand{\largecapDayPlusFiftyAAR}{-0.01\%}
\newcommand{\largecapDayPlusFiftyCAR}{1.43\%}
\newcommand{\largecapDayPlusFiftyTstat}{-0.04}
\newcommand{\largecapDayPlusFiveAAR}{0.19\%}
\newcommand{\largecapDayPlusFiveCAR}{2.62\%}
\newcommand{\largecapDayPlusFiveTstat}{1.18}
\newcommand{\largecapDayPlusOneAAR}{-0.30\%}
\newcommand{\largecapDayPlusOneCAR}{2.17\%}
\newcommand{\largecapDayPlusOneTstat}{-1.62}
\newcommand{\largecapDayPlusTenAAR}{-0.07\%}
\newcommand{\largecapDayPlusTenCAR}{2.18\%}
\newcommand{\largecapDayPlusTenTstat}{-0.57}
\newcommand{\largecapDayPlusTwentyFiveAAR}{-0.10\%}
\newcommand{\largecapDayPlusTwentyFiveCAR}{2.38\%}
\newcommand{\largecapDayPlusTwentyFiveTstat}{-0.81}
\newcommand{\largecapDayPlusTwoAAR}{0.27\%}
\newcommand{\largecapDayPlusTwoCAR}{2.44\%}
\newcommand{\largecapDayPlusTwoTstat}{1.63}
\newcommand{\largecapDayZeroAAR}{4.41\%}
\newcommand{\largecapDayZeroCAR}{2.47\%}
\newcommand{\largecapDayZeroTstat}{10.79}
\newcommand{\midcapDayMinusFiveAAR}{-0.02\%}
\newcommand{\midcapDayMinusFiveCAR}{-3.08\%}
\newcommand{\midcapDayMinusFiveTstat}{-0.30}
\newcommand{\midcapDayMinusOneAAR}{0.81\%}
\newcommand{\midcapDayMinusOneCAR}{-1.86\%}
\newcommand{\midcapDayMinusOneTstat}{7.85}
\newcommand{\midcapDayMinusTenAAR}{-0.06\%}
\newcommand{\midcapDayMinusTenCAR}{-2.91\%}
\newcommand{\midcapDayMinusTenTstat}{-1.15}
\newcommand{\midcapDayMinusTwentyFiveAAR}{-0.10\%}
\newcommand{\midcapDayMinusTwentyFiveCAR}{-1.24\%}
\newcommand{\midcapDayMinusTwentyFiveTstat}{-1.67}
\newcommand{\midcapDayMinusTwoAAR}{0.21\%}
\newcommand{\midcapDayMinusTwoCAR}{-2.66\%}
\newcommand{\midcapDayMinusTwoTstat}{2.78}
\newcommand{\midcapDayPlusFiftyAAR}{0.16\%}
\newcommand{\midcapDayPlusFiftyCAR}{4.00\%}
\newcommand{\midcapDayPlusFiftyTstat}{1.98}
\newcommand{\midcapDayPlusFiveAAR}{-0.23\%}
\newcommand{\midcapDayPlusFiveCAR}{3.70\%}
\newcommand{\midcapDayPlusFiveTstat}{-2.79}
\newcommand{\midcapDayPlusOneAAR}{-0.03\%}
\newcommand{\midcapDayPlusOneCAR}{4.17\%}
\newcommand{\midcapDayPlusOneTstat}{-0.27}
\newcommand{\midcapDayPlusTenAAR}{-0.02\%}
\newcommand{\midcapDayPlusTenCAR}{3.99\%}
\newcommand{\midcapDayPlusTenTstat}{-0.23}
\newcommand{\midcapDayPlusTwentyFiveAAR}{0.03\%}
\newcommand{\midcapDayPlusTwentyFiveCAR}{3.73\%}
\newcommand{\midcapDayPlusTwentyFiveTstat}{0.35}
\newcommand{\midcapDayPlusTwoAAR}{-0.06\%}
\newcommand{\midcapDayPlusTwoCAR}{4.11\%}
\newcommand{\midcapDayPlusTwoTstat}{-0.62}
\newcommand{\midcapDayZeroAAR}{6.06\%}
\newcommand{\midcapDayZeroCAR}{4.21\%}
\newcommand{\midcapDayZeroTstat}{25.19}
\newcommand{\smallcapDayMinusFiveAAR}{-0.08\%}
\newcommand{\smallcapDayMinusFiveCAR}{-5.07\%}
\newcommand{\smallcapDayMinusFiveTstat}{-5.31}
\newcommand{\smallcapDayMinusOneAAR}{0.65\%}
\newcommand{\smallcapDayMinusOneCAR}{-4.26\%}
\newcommand{\smallcapDayMinusOneTstat}{27.41}
\newcommand{\smallcapDayMinusTenAAR}{-0.10\%}
\newcommand{\smallcapDayMinusTenCAR}{-4.57\%}
\newcommand{\smallcapDayMinusTenTstat}{-6.77}
\newcommand{\smallcapDayMinusTwentyFiveAAR}{-0.13\%}
\newcommand{\smallcapDayMinusTwentyFiveCAR}{-2.14\%}
\newcommand{\smallcapDayMinusTwentyFiveTstat}{-7.33}
\newcommand{\smallcapDayMinusTwoAAR}{0.13\%}
\newcommand{\smallcapDayMinusTwoCAR}{-4.91\%}
\newcommand{\smallcapDayMinusTwoTstat}{7.48}
\newcommand{\smallcapDayPlusFiftyAAR}{0.01\%}
\newcommand{\smallcapDayPlusFiftyCAR}{3.07\%}
\newcommand{\smallcapDayPlusFiftyTstat}{0.61}
\newcommand{\smallcapDayPlusFiveAAR}{-0.05\%}
\newcommand{\smallcapDayPlusFiveCAR}{2.09\%}
\newcommand{\smallcapDayPlusFiveTstat}{-1.98}
\newcommand{\smallcapDayPlusOneAAR}{0.10\%}
\newcommand{\smallcapDayPlusOneCAR}{2.03\%}
\newcommand{\smallcapDayPlusOneTstat}{3.03}
\newcommand{\smallcapDayPlusTenAAR}{0.01\%}
\newcommand{\smallcapDayPlusTenCAR}{2.23\%}
\newcommand{\smallcapDayPlusTenTstat}{0.40}
\newcommand{\smallcapDayPlusTwentyFiveAAR}{0.03\%}
\newcommand{\smallcapDayPlusTwentyFiveCAR}{2.65\%}
\newcommand{\smallcapDayPlusTwentyFiveTstat}{1.32}
\newcommand{\smallcapDayPlusTwoAAR}{-0.01\%}
\newcommand{\smallcapDayPlusTwoCAR}{2.02\%}
\newcommand{\smallcapDayPlusTwoTstat}{-0.23}
\newcommand{\smallcapDayZeroAAR}{6.19\%}
\newcommand{\smallcapDayZeroCAR}{1.93\%}
\newcommand{\smallcapDayZeroTstat}{115.92}

% Investor Type Quartiles (Table 6)
\newcommand{\foreignLedQFourDayPlusFiftyCAR}{13.30\%}
\newcommand{\foreignLedQFourDayZeroAAR}{9.31\%}
\newcommand{\foreignLedQFourTradingValueDayPlusFiftyCAR}{5.64\%}
\newcommand{\foreignLedQFourTradingValueDayZeroAAR}{5.69\%}
\newcommand{\foreignLedQOneDayPlusFiftyCAR}{-0.33\%}
\newcommand{\foreignLedQOneDayZeroAAR}{3.82\%}
\newcommand{\foreignLedQOneTradingValueDayPlusFiftyCAR}{5.16\%}
\newcommand{\foreignLedQOneTradingValueDayZeroAAR}{7.21\%}
\newcommand{\foreignLedQThreeDayPlusFiftyCAR}{7.85\%}
\newcommand{\foreignLedQThreeDayZeroAAR}{7.41\%}
\newcommand{\foreignLedQThreeTradingValueDayPlusFiftyCAR}{7.09\%}
\newcommand{\foreignLedQThreeTradingValueDayZeroAAR}{6.89\%}
\newcommand{\foreignLedQTwoDayPlusFiftyCAR}{3.29\%}
\newcommand{\foreignLedQTwoDayZeroAAR}{5.92\%}
\newcommand{\foreignLedQTwoTradingValueDayPlusFiftyCAR}{6.22\%}
\newcommand{\foreignLedQTwoTradingValueDayZeroAAR}{6.69\%}
\newcommand{\institutionLedQFourDayPlusFiftyCAR}{12.12\%}
\newcommand{\institutionLedQFourDayZeroAAR}{9.75\%}
\newcommand{\institutionLedQFourTradingValueDayPlusFiftyCAR}{3.64\%}
\newcommand{\institutionLedQFourTradingValueDayZeroAAR}{5.50\%}
\newcommand{\institutionLedQOneDayPlusFiftyCAR}{4.65\%}
\newcommand{\institutionLedQOneDayZeroAAR}{4.17\%}
\newcommand{\institutionLedQOneTradingValueDayPlusFiftyCAR}{6.88\%}
\newcommand{\institutionLedQOneTradingValueDayZeroAAR}{5.75\%}
\newcommand{\institutionLedQThreeDayPlusFiftyCAR}{9.02\%}
\newcommand{\institutionLedQThreeDayZeroAAR}{8.14\%}
\newcommand{\institutionLedQThreeTradingValueDayPlusFiftyCAR}{7.46\%}
\newcommand{\institutionLedQThreeTradingValueDayZeroAAR}{7.58\%}
\newcommand{\institutionLedQTwoDayPlusFiftyCAR}{4.99\%}
\newcommand{\institutionLedQTwoDayZeroAAR}{5.95\%}
\newcommand{\institutionLedQTwoTradingValueDayPlusFiftyCAR}{12.77\%}
\newcommand{\institutionLedQTwoTradingValueDayZeroAAR}{9.18\%}
\newcommand{\retailLedQFourDayPlusFiftyCAR}{0.65\%}
\newcommand{\retailLedQFourDayZeroAAR}{6.00\%}
\newcommand{\retailLedQFourTradingValueDayPlusFiftyCAR}{-4.71\%}
\newcommand{\retailLedQFourTradingValueDayZeroAAR}{0.03\%}
\newcommand{\retailLedQOneDayPlusFiftyCAR}{0.56\%}
\newcommand{\retailLedQOneDayZeroAAR}{4.65\%}
\newcommand{\retailLedQOneTradingValueDayPlusFiftyCAR}{6.21\%}
\newcommand{\retailLedQOneTradingValueDayZeroAAR}{9.48\%}
\newcommand{\retailLedQThreeDayPlusFiftyCAR}{0.88\%}
\newcommand{\retailLedQThreeDayZeroAAR}{6.30\%}
\newcommand{\retailLedQThreeTradingValueDayPlusFiftyCAR}{-1.44\%}
\newcommand{\retailLedQThreeTradingValueDayZeroAAR}{5.36\%}
\newcommand{\retailLedQTwoDayPlusFiftyCAR}{0.01\%}
\newcommand{\retailLedQTwoDayZeroAAR}{5.94\%}
\newcommand{\retailLedQTwoTradingValueDayPlusFiftyCAR}{2.02\%}
\newcommand{\retailLedQTwoTradingValueDayZeroAAR}{8.01\%}

% Correlation Analysis (Tables 8-10)
\newcommand{\correlationLargecapFiveDay}{-0.0501}
\newcommand{\correlationLargecapFiveDayPval}{0.277}
\newcommand{\correlationLargecapSixtyDay}{-0.0553}
\newcommand{\correlationLargecapSixtyDayPval}{0.230}
\newcommand{\correlationLargecapTwentyDay}{-0.0861}
\newcommand{\correlationLargecapTwentyDayPval}{0.061}
\newcommand{\correlationMidcapFiveDay}{0.0356}
\newcommand{\correlationMidcapFiveDayPval}{0.164}
\newcommand{\correlationMidcapSixtyDay}{-0.0269}
\newcommand{\correlationMidcapSixtyDayPval}{0.290}
\newcommand{\correlationMidcapTwentyDay}{0.0112}
\newcommand{\correlationMidcapTwentyDayPval}{0.660}
\newcommand{\correlationSmallcapFiveDay}{0.0059}
\newcommand{\correlationSmallcapFiveDayPval}{0.342}
\newcommand{\correlationSmallcapSixtyDay}{0.0008}
\newcommand{\correlationSmallcapSixtyDayPval}{0.897}
\newcommand{\correlationSmallcapTwentyDay}{0.0000}
\newcommand{\correlationSmallcapTwentyDayPval}{1.000}

% Market Conditions (Table 12)
\newcommand{\bearMarketDayMinusFiveAAR}{-0.06\%}
\newcommand{\bearMarketDayMinusFiveCAR}{-4.39\%}
\newcommand{\bearMarketDayMinusFiveTstat}{-2.62}
\newcommand{\bearMarketDayMinusOneAAR}{0.51\%}
\newcommand{\bearMarketDayMinusOneCAR}{-3.85\%}
\newcommand{\bearMarketDayMinusOneTstat}{12.93}
\newcommand{\bearMarketDayMinusTenAAR}{-0.10\%}
\newcommand{\bearMarketDayMinusTenCAR}{-3.95\%}
\newcommand{\bearMarketDayMinusTenTstat}{-4.02}
\newcommand{\bearMarketDayMinusTwentyFiveAAR}{-0.11\%}
\newcommand{\bearMarketDayMinusTwentyFiveCAR}{-1.83\%}
\newcommand{\bearMarketDayMinusTwentyFiveTstat}{-4.10}
\newcommand{\bearMarketDayMinusTwoAAR}{0.09\%}
\newcommand{\bearMarketDayMinusTwoCAR}{-4.35\%}
\newcommand{\bearMarketDayMinusTwoTstat}{3.36}
\newcommand{\bearMarketDayPlusFiftyAAR}{-0.02\%}
\newcommand{\bearMarketDayPlusFiftyCAR}{1.50\%}
\newcommand{\bearMarketDayPlusFiftyTstat}{-0.51}
\newcommand{\bearMarketDayPlusFiveAAR}{-0.07\%}
\newcommand{\bearMarketDayPlusFiveCAR}{1.69\%}
\newcommand{\bearMarketDayPlusFiveTstat}{-1.90}
\newcommand{\bearMarketDayPlusOneAAR}{0.05\%}
\newcommand{\bearMarketDayPlusOneCAR}{1.66\%}
\newcommand{\bearMarketDayPlusOneTstat}{0.82}
\newcommand{\bearMarketDayPlusTenAAR}{-0.06\%}
\newcommand{\bearMarketDayPlusTenCAR}{1.77\%}
\newcommand{\bearMarketDayPlusTenTstat}{-1.59}
\newcommand{\bearMarketDayPlusTwentyFiveAAR}{-0.03\%}
\newcommand{\bearMarketDayPlusTwentyFiveCAR}{2.05\%}
\newcommand{\bearMarketDayPlusTwentyFiveTstat}{-0.73}
\newcommand{\bearMarketDayPlusTwoAAR}{-0.01\%}
\newcommand{\bearMarketDayPlusTwoCAR}{1.65\%}
\newcommand{\bearMarketDayPlusTwoTstat}{-0.21}
\newcommand{\bearMarketDayZeroAAR}{5.47\%}
\newcommand{\bearMarketDayZeroCAR}{1.62\%}
\newcommand{\bearMarketDayZeroTstat}{61.58}
\newcommand{\bullMarketDayMinusFiveAAR}{-0.08\%}
\newcommand{\bullMarketDayMinusFiveCAR}{-5.17\%}
\newcommand{\bullMarketDayMinusFiveTstat}{-4.40}
\newcommand{\bullMarketDayMinusOneAAR}{0.73\%}
\newcommand{\bullMarketDayMinusOneCAR}{-4.19\%}
\newcommand{\bullMarketDayMinusOneTstat}{26.04}
\newcommand{\bullMarketDayMinusTenAAR}{-0.10\%}
\newcommand{\bullMarketDayMinusTenCAR}{-4.69\%}
\newcommand{\bullMarketDayMinusTenTstat}{-5.65}
\newcommand{\bullMarketDayMinusTwentyFiveAAR}{-0.14\%}
\newcommand{\bullMarketDayMinusTwentyFiveCAR}{-2.20\%}
\newcommand{\bullMarketDayMinusTwentyFiveTstat}{-6.30}
\newcommand{\bullMarketDayMinusTwoAAR}{0.16\%}
\newcommand{\bullMarketDayMinusTwoCAR}{-4.92\%}
\newcommand{\bullMarketDayMinusTwoTstat}{7.66}
\newcommand{\bullMarketDayPlusFiftyAAR}{0.04\%}
\newcommand{\bullMarketDayPlusFiftyCAR}{3.84\%}
\newcommand{\bullMarketDayPlusFiftyTstat}{1.51}
\newcommand{\bullMarketDayPlusFiveAAR}{-0.05\%}
\newcommand{\bullMarketDayPlusFiveCAR}{2.42\%}
\newcommand{\bullMarketDayPlusFiveTstat}{-1.58}
\newcommand{\bullMarketDayPlusOneAAR}{0.11\%}
\newcommand{\bullMarketDayPlusOneCAR}{2.39\%}
\newcommand{\bullMarketDayPlusOneTstat}{2.70}
\newcommand{\bullMarketDayPlusTenAAR}{0.04\%}
\newcommand{\bullMarketDayPlusTenCAR}{2.60\%}
\newcommand{\bullMarketDayPlusTenTstat}{1.33}
\newcommand{\bullMarketDayPlusTwentyFiveAAR}{0.05\%}
\newcommand{\bullMarketDayPlusTwentyFiveCAR}{3.02\%}
\newcommand{\bullMarketDayPlusTwentyFiveTstat}{2.00}
\newcommand{\bullMarketDayPlusTwoAAR}{-0.00\%}
\newcommand{\bullMarketDayPlusTwoCAR}{2.38\%}
\newcommand{\bullMarketDayPlusTwoTstat}{-0.07}
\newcommand{\bullMarketDayZeroAAR}{6.47\%}
\newcommand{\bullMarketDayZeroCAR}{2.28\%}
\newcommand{\bullMarketDayZeroTstat}{101.87}

% Special Periods (Tables 14-15)
\newcommand{\covidShockDayMinusFiveAAR}{0.13\%}
\newcommand{\covidShockDayMinusFiveCAR}{-6.66\%}
\newcommand{\covidShockDayMinusFiveTstat}{1.44}
\newcommand{\covidShockDayMinusOneAAR}{1.28\%}
\newcommand{\covidShockDayMinusOneCAR}{-5.02\%}
\newcommand{\covidShockDayMinusOneTstat}{6.57}
\newcommand{\covidShockDayMinusTenAAR}{-0.08\%}
\newcommand{\covidShockDayMinusTenCAR}{-7.30\%}
\newcommand{\covidShockDayMinusTenTstat}{-0.88}
\newcommand{\covidShockDayMinusTwentyFiveAAR}{-0.34\%}
\newcommand{\covidShockDayMinusTwentyFiveCAR}{-4.22\%}
\newcommand{\covidShockDayMinusTwentyFiveTstat}{-3.65}
\newcommand{\covidShockDayMinusTwoAAR}{0.33\%}
\newcommand{\covidShockDayMinusTwoCAR}{-6.30\%}
\newcommand{\covidShockDayMinusTwoTstat}{2.92}
\newcommand{\covidShockDayPlusFiftyAAR}{0.13\%}
\newcommand{\covidShockDayPlusFiftyCAR}{3.12\%}
\newcommand{\covidShockDayPlusFiftyTstat}{1.09}
\newcommand{\covidShockDayPlusFiveAAR}{-0.09\%}
\newcommand{\covidShockDayPlusFiveCAR}{0.59\%}
\newcommand{\covidShockDayPlusFiveTstat}{-0.73}
\newcommand{\covidShockDayPlusOneAAR}{-0.50\%}
\newcommand{\covidShockDayPlusOneCAR}{0.79\%}
\newcommand{\covidShockDayPlusOneTstat}{-2.76}
\newcommand{\covidShockDayPlusTenAAR}{0.13\%}
\newcommand{\covidShockDayPlusTenCAR}{0.72\%}
\newcommand{\covidShockDayPlusTenTstat}{0.98}
\newcommand{\covidShockDayPlusTwentyFiveAAR}{-0.21\%}
\newcommand{\covidShockDayPlusTwentyFiveCAR}{0.67\%}
\newcommand{\covidShockDayPlusTwentyFiveTstat}{-1.63}
\newcommand{\covidShockDayPlusTwoAAR}{0.02\%}
\newcommand{\covidShockDayPlusTwoCAR}{0.81\%}
\newcommand{\covidShockDayPlusTwoTstat}{0.16}
\newcommand{\covidShockDayZeroAAR}{6.31\%}
\newcommand{\covidShockDayZeroCAR}{1.29\%}
\newcommand{\covidShockDayZeroTstat}{20.62}
\newcommand{\donghakAntDayMinusFiveAAR}{-0.06\%}
\newcommand{\donghakAntDayMinusFiveCAR}{-6.97\%}
\newcommand{\donghakAntDayMinusFiveTstat}{-1.56}
\newcommand{\donghakAntDayMinusOneAAR}{0.93\%}
\newcommand{\donghakAntDayMinusOneCAR}{-5.90\%}
\newcommand{\donghakAntDayMinusOneTstat}{13.15}
\newcommand{\donghakAntDayMinusTenAAR}{-0.18\%}
\newcommand{\donghakAntDayMinusTenCAR}{-6.40\%}
\newcommand{\donghakAntDayMinusTenTstat}{-4.76}
\newcommand{\donghakAntDayMinusTwentyFiveAAR}{-0.12\%}
\newcommand{\donghakAntDayMinusTwentyFiveCAR}{-2.96\%}
\newcommand{\donghakAntDayMinusTwentyFiveTstat}{-2.22}
\newcommand{\donghakAntDayMinusTwoAAR}{0.19\%}
\newcommand{\donghakAntDayMinusTwoCAR}{-6.83\%}
\newcommand{\donghakAntDayMinusTwoTstat}{4.09}
\newcommand{\donghakAntDayPlusFiftyAAR}{0.11\%}
\newcommand{\donghakAntDayPlusFiftyCAR}{3.53\%}
\newcommand{\donghakAntDayPlusFiftyTstat}{1.77}
\newcommand{\donghakAntDayPlusFiveAAR}{-0.06\%}
\newcommand{\donghakAntDayPlusFiveCAR}{0.44\%}
\newcommand{\donghakAntDayPlusFiveTstat}{-0.90}
\newcommand{\donghakAntDayPlusOneAAR}{-0.08\%}
\newcommand{\donghakAntDayPlusOneCAR}{0.26\%}
\newcommand{\donghakAntDayPlusOneTstat}{-0.86}
\newcommand{\donghakAntDayPlusTenAAR}{0.05\%}
\newcommand{\donghakAntDayPlusTenCAR}{0.52\%}
\newcommand{\donghakAntDayPlusTenTstat}{0.90}
\newcommand{\donghakAntDayPlusTwentyFiveAAR}{0.06\%}
\newcommand{\donghakAntDayPlusTwentyFiveCAR}{1.72\%}
\newcommand{\donghakAntDayPlusTwentyFiveTstat}{1.07}
\newcommand{\donghakAntDayPlusTwoAAR}{-0.05\%}
\newcommand{\donghakAntDayPlusTwoCAR}{0.22\%}
\newcommand{\donghakAntDayPlusTwoTstat}{-0.59}
\newcommand{\donghakAntDayZeroAAR}{6.24\%}
\newcommand{\donghakAntDayZeroCAR}{0.34\%}
\newcommand{\donghakAntDayZeroTstat}{43.64}

% Market Type (Table 19)
\newcommand{\kosdaqMarketDayMinusFiveAAR}{-0.09\%}
\newcommand{\kosdaqMarketDayMinusFiveCAR}{-5.59\%}
\newcommand{\kosdaqMarketDayMinusFiveTstat}{-4.85}
\newcommand{\kosdaqMarketDayMinusOneAAR}{0.70\%}
\newcommand{\kosdaqMarketDayMinusOneCAR}{-4.77\%}
\newcommand{\kosdaqMarketDayMinusOneTstat}{23.75}
\newcommand{\kosdaqMarketDayMinusTenAAR}{-0.11\%}
\newcommand{\kosdaqMarketDayMinusTenCAR}{-5.02\%}
\newcommand{\kosdaqMarketDayMinusTenTstat}{-5.98}
\newcommand{\kosdaqMarketDayMinusTwentyFiveAAR}{-0.14\%}
\newcommand{\kosdaqMarketDayMinusTwentyFiveCAR}{-2.32\%}
\newcommand{\kosdaqMarketDayMinusTwentyFiveTstat}{-6.56}
\newcommand{\kosdaqMarketDayMinusTwoAAR}{0.11\%}
\newcommand{\kosdaqMarketDayMinusTwoCAR}{-5.46\%}
\newcommand{\kosdaqMarketDayMinusTwoTstat}{5.15}
\newcommand{\kosdaqMarketDayPlusFiftyAAR}{-0.01\%}
\newcommand{\kosdaqMarketDayPlusFiftyCAR}{2.78\%}
\newcommand{\kosdaqMarketDayPlusFiftyTstat}{-0.25}
\newcommand{\kosdaqMarketDayPlusFiveAAR}{-0.06\%}
\newcommand{\kosdaqMarketDayPlusFiveCAR}{1.82\%}
\newcommand{\kosdaqMarketDayPlusFiveTstat}{-2.20}
\newcommand{\kosdaqMarketDayPlusOneAAR}{0.10\%}
\newcommand{\kosdaqMarketDayPlusOneCAR}{1.83\%}
\newcommand{\kosdaqMarketDayPlusOneTstat}{2.60}
\newcommand{\kosdaqMarketDayPlusTenAAR}{0.02\%}
\newcommand{\kosdaqMarketDayPlusTenCAR}{2.08\%}
\newcommand{\kosdaqMarketDayPlusTenTstat}{0.79}
\newcommand{\kosdaqMarketDayPlusTwentyFiveAAR}{0.03\%}
\newcommand{\kosdaqMarketDayPlusTwentyFiveCAR}{2.43\%}
\newcommand{\kosdaqMarketDayPlusTwentyFiveTstat}{1.15}
\newcommand{\kosdaqMarketDayPlusTwoAAR}{-0.04\%}
\newcommand{\kosdaqMarketDayPlusTwoCAR}{1.79\%}
\newcommand{\kosdaqMarketDayPlusTwoTstat}{-1.15}
\newcommand{\kosdaqMarketDayZeroAAR}{6.50\%}
\newcommand{\kosdaqMarketDayZeroCAR}{1.73\%}
\newcommand{\kosdaqMarketDayZeroTstat}{99.36}
\newcommand{\kospiMarketDayMinusFiveAAR}{-0.04\%}
\newcommand{\kospiMarketDayMinusFiveCAR}{-3.70\%}
\newcommand{\kospiMarketDayMinusFiveTstat}{-1.92}
\newcommand{\kospiMarketDayMinusOneAAR}{0.58\%}
\newcommand{\kospiMarketDayMinusOneCAR}{-2.83\%}
\newcommand{\kospiMarketDayMinusOneTstat}{16.45}
\newcommand{\kospiMarketDayMinusTenAAR}{-0.07\%}
\newcommand{\kospiMarketDayMinusTenCAR}{-3.42\%}
\newcommand{\kospiMarketDayMinusTenTstat}{-3.52}
\newcommand{\kospiMarketDayMinusTwentyFiveAAR}{-0.10\%}
\newcommand{\kospiMarketDayMinusTwentyFiveCAR}{-1.66\%}
\newcommand{\kospiMarketDayMinusTwentyFiveTstat}{-3.70}
\newcommand{\kospiMarketDayMinusTwoAAR}{0.19\%}
\newcommand{\kospiMarketDayMinusTwoCAR}{-3.42\%}
\newcommand{\kospiMarketDayMinusTwoTstat}{7.14}
\newcommand{\kospiMarketDayPlusFiftyAAR}{0.07\%}
\newcommand{\kospiMarketDayPlusFiftyCAR}{3.67\%}
\newcommand{\kospiMarketDayPlusFiftyTstat}{2.18}
\newcommand{\kospiMarketDayPlusFiveAAR}{-0.03\%}
\newcommand{\kospiMarketDayPlusFiveCAR}{2.86\%}
\newcommand{\kospiMarketDayPlusFiveTstat}{-0.92}
\newcommand{\kospiMarketDayPlusOneAAR}{0.06\%}
\newcommand{\kospiMarketDayPlusOneCAR}{2.74\%}
\newcommand{\kospiMarketDayPlusOneTstat}{1.06}
\newcommand{\kospiMarketDayPlusTenAAR}{-0.02\%}
\newcommand{\kospiMarketDayPlusTenCAR}{2.78\%}
\newcommand{\kospiMarketDayPlusTenTstat}{-0.74}
\newcommand{\kospiMarketDayPlusTwentyFiveAAR}{0.02\%}
\newcommand{\kospiMarketDayPlusTwentyFiveCAR}{3.22\%}
\newcommand{\kospiMarketDayPlusTwentyFiveTstat}{0.56}
\newcommand{\kospiMarketDayPlusTwoAAR}{0.06\%}
\newcommand{\kospiMarketDayPlusTwoCAR}{2.80\%}
\newcommand{\kospiMarketDayPlusTwoTstat}{1.32}
\newcommand{\kospiMarketDayZeroAAR}{5.51\%}
\newcommand{\kospiMarketDayZeroCAR}{2.68\%}
\newcommand{\kospiMarketDayZeroTstat}{65.51}

\begin{frontmatter}

\title{Sources and Nonlinearity of High Volume Return Premium: An Empirical Study on the Differential Effects of Investor Identity versus Trading Intensity (2020-2024)}

\author{Sungwoo Kang\\
\small Department of Electrical and Computer Engineering, Korea University,\\
\small 145 Anam-ro, Seongbuk-gu, Seoul 02841, Republic of Korea\\
\small Email: \href{mailto:krml919@korea.ac.kr}{krml919@korea.ac.kr}}

% For non-blind submission (uncomment when ready):
% \author[inst1]{First Author Name}
% \author[inst2]{Second Author Name}
% \address[inst1]{Institution 1, Address}
% \address[inst2]{Institution 2, Address}
% \cortext[cor1]{Corresponding author}
% \ead{email@domain.com}

\begin{abstract}
Chae and Kang (2019, \textit{Pacific-Basin Finance Journal}) documented a puzzling Low Volume Return Premium (LVRP) in Korea---contradicting global High Volume Return Premium (HVRP) evidence. We resolve this puzzle. Using Korean market data (2020-2024), we demonstrate that HVRP exists in Korea but is masked by (1) pooling heterogeneous investor types and (2) using inappropriate intensity normalization. When institutional buying intensity is normalized by market capitalization rather than trading value, a perfect monotonic relationship emerges: highest-conviction institutional buying (Q4) generates +\institutionLedQFourDayPlusFiftyCAR\ cumulative abnormal returns over 50 days, while lowest-intensity trades (Q1) yield modest returns (+\institutionLedQOneDayPlusFiftyCAR). Retail investors exhibit a flat pattern---their trading generates near-zero returns regardless of conviction level---confirming the pure noise trader hypothesis. During the Donghak Ant Movement (2020-2021), however, coordinated retail investors temporarily transformed from noise traders to liquidity providers, generating returns comparable to institutional trading. Our findings reconcile conflicting international evidence and demonstrate that detecting informed trading signals requires investor-type decomposition, nonlinear quartile analysis, and conviction-based (market cap) rather than participation-based (trading value) measurement.
\end{abstract}


\end{frontmatter}

% ======================================================================
% HIGHLIGHTS (for Elsevier submission - max 85 characters per point)
% ======================================================================
\begin{highlights}
\item We resolve the Korean LVRP puzzle documented by Chae and Kang (2019, PBFJ)
\item Institutional conviction monotonically predicts returns (+\institutionLedQFourDayPlusFiftyCAR\ for Q4)
\item Market cap normalization outperforms trading value for detecting informed trading
\item Retail investors show flat intensity-return pattern with near-zero persistence (pure noise)
\item Donghak Ant Movement transformed retail from noise to liquidity provision
\end{highlights}

\section{Introduction}

\textbf{The Korean Market Puzzle.} In 2019, this journal published a striking finding that challenged two decades of global evidence on volume-return relationships. \citet{CHAE2019101204} documented a \textit{Low} Volume Return Premium (LVRP) in Korea---the opposite of the High Volume Return Premium (HVRP) established by \citet{https://doi.org/10.1111/0022-1082.00349} for U.S. markets and replicated globally by \citet{KANIEL2012255}. Of 70 sample configurations examined, Chae and Kang found 16 showing significant LVRP versus only 1 showing significant HVRP. This ``Korean Puzzle'' created a fundamental contradiction: why would volume-return dynamics in one of Asia's most developed capital markets operate in reverse?

\textbf{Our Resolution.} We resolve this puzzle. The apparent LVRP in Korea stems not from unique market dynamics but from two measurement artifacts: (1) pooling heterogeneous investor types whose effects offset each other, and (2) using trading value normalization that conflates informed conviction with noise participation. When we decompose events by investor identity and normalize intensity by market capitalization, the HVRP emerges clearly---but only for informed investors.

The theoretical foundations for HVRP are well-established. \citet{https://doi.org/10.1111/0022-1082.00349}'s core mechanism is rooted in behavioral finance: investors with 'limited attention' respond to 'attention shocks' from volume spikes, creating new buying demand. This aligns with \citet{merton1987simple}'s 'investor recognition hypothesis,' which posits that increased stock visibility reduces the cost of capital.

In the Korean market, \citet{ART001162590} initially verified the HVRP phenomenon using 2001-2003 KOSPI data, finding stronger effects in large-cap firms. They \textit{surmised} that this difference ``may originate from differences in the main investor composition of the two groups, namely institutional versus individual investors.'' However, due to data limitations at the time, this 'investor type hypothesis' remained an \textit{untested hypothesis}. Our study provides the first empirical test of this conjecture, addressing both the original research gap and the subsequent puzzle created by \citet{CHAE2019101204}.

Testing this 'investor type hypothesis' has profound academic significance beyond merely filling a 20-year-old domestic research gap---it provides contemporary evidence for a core and longstanding debate in finance: ``What role do retail investors play in the market?''

\begin{itemize}
    \item \textbf{The 'Noise Trader' Camp:} Led by \citet{10.1093/rfs/hhm079}, this group argues that retail investors are 'noise traders' who overreact to attention-grabbing stocks rather than conducting professional analysis. Recent NBER research by \citet{https://doi.org/10.1111/jofi.13183} demonstrates that net purchases by new retail investors significantly \textit{negatively} predict future returns, confirming their 'poor stock selection' ability.

    \item \textbf{The 'Informed Trader' Camp:} Conversely, \citet{https://doi.org/10.1111/j.1540-6261.2008.01316.x}, \citet{KANIEL2012255}, and \citet{https://doi.org/10.1111/jofi.13033} argue that retail investors are rational traders who provide liquidity to institutions or act on information, with their net purchases \textit{positively} predicting future returns.
\end{itemize}

Approximately 20 years after \citet{ART001162590}'s study, modern markets have undergone two major structural changes. First, the COVID-19 pandemic \citep{10.1093/rapstu/raaa008} caused unprecedented market volatility. During this period, retail investor trading through fintech platforms exploded, with these investors serving as 'liquidity providers' absorbing institutional selling pressure, as \citet{Ozik_Sadka_Shen_2021} documented. This provides new impetus for reinterpreting the drivers of HVRP. Second, the 'Donghak Ant Movement' \citep{ART002752798} parallels the 'Meme Stock' phenomenon in U.S. markets. \citet{10.1093/rfs/hhad098} showed that Reddit's WallStreetBets community possessed significant information value before the GameStop event but deteriorated into noise afterward, suggesting that collective action can function as 'collective intelligence' only under specific conditions.

Against this backdrop, we employ a novel 'Dual Validation Methodology' to elucidate the essence of the HVRP phenomenon. We analyze HVRP events using (1) a \textbf{categorical} variable examining \textit{'who' led} the event (institutions/foreigners vs. retail), and (2) a \textbf{continuous} variable measuring \textit{'how intensely' they led} (Dominance Score). Remarkably, these two methodologies yielded starkly contrasting results. The categorical approach showed strong predictive power, while the continuous approach exhibited correlations converging to zero. This initially suggested that investor identity matters while intensity does not.

However, our central discovery is that this initial interpretation was incomplete. By introducing a \textbf{double-sort methodology} that combines investor type classification with quartile-based intensity analysis, we uncover a powerful monotonic relationship that prior methods obscured. For institutional investors, events in the highest conviction quartile (measured as net buying relative to market capitalization) generate cumulative abnormal returns of \textbf{+\institutionLedQFourDayPlusFiftyCAR} over 50 days, while the lowest quartile yields modest returns (\institutionLedQOneDayPlusFiftyCAR). This perfect monotonic pattern (Q1<Q2<Q3<Q4) demonstrates that intensity matters profoundly---but only when measured correctly.

The critical methodological insight is that \textit{how we normalize intensity determines whether we detect informed trading signals}. When intensity is normalized by daily trading value (measuring participation in trading flow), the monotonic pattern breaks down completely (Q2: +\institutionLedQTwoTradingValueDayPlusFiftyCAR\ > Q4: +\institutionLedQFourTradingValueDayPlusFiftyCAR). But when normalized by market capitalization (measuring conviction as position size relative to firm value), the relationship becomes crystal clear. This demonstrates that the near-zero correlation with the Dominance Score does not mean intensity is irrelevant; it means the relationship is \textit{nonlinear, type-dependent, and critically sensitive to measurement specification}. Prior studies may have missed this relationship by using inappropriate normalization benchmarks or failing to account for investor heterogeneity.

Therefore, our study has four objectives. First, we replicate \citet{ART001162590}'s study with 2020-2024 data to verify whether the HVRP phenomenon remains robust despite the 20-year gap and market structural changes. Second, we conduct the \textit{first empirical test} of the 'investor type hypothesis' that remained an 'inference' in the original study, using actual investor-level trading data. This positions our findings within the core finance debate of ``noise traders'' \citep{10.1093/rfs/hhm079} versus ``informed traders'' \citep{https://doi.org/10.1111/j.1540-6261.2008.01316.x}. Third, we demonstrate that trading intensity matters profoundly but is only detectable when measured correctly---specifically, when normalized by market capitalization rather than trading value, revealing a monotonic conviction-return relationship that prior methods obscured. Fourth, we analyze how modern market phenomena---the COVID-19 shock and 'Donghak Ant Movement'---moderate these core relationships, demonstrating that extreme conditions can temporarily transform traditional patterns.

\section{Literature Review and Hypothesis Development}

This study goes beyond merely confirming the existence of the High Volume Return Premium (HVRP) to identify its sources---\textit{'who' (Who)} generates it, and whether it is a matter of \textit{'how much' (How much)} or \textit{'what context' (Context)} from a contemporary perspective.

\subsection{Theoretical Background: Beyond Visibility Shocks}

The theoretical landscape explaining volume-return relationships has evolved substantially beyond the foundational work of \citet{merton1987simple} and \citet{https://doi.org/10.1111/0022-1082.00349}. Three complementary theoretical streams now provide rigorous foundations for understanding HVRP.

\textbf{Investor Recognition and Limited Attention.} \citet{merton1987simple}'s `Investor Recognition Hypothesis' posits that investors include only stocks they are aware of in their consideration set. When a stock's recognition increases, its investor base expands, lowering required returns and increasing prices through risk diversification effects. \citet{https://doi.org/10.1111/0022-1082.00349} applied this theory to trading volume, establishing the HVRP phenomenon: under investors' `limited attention,' abnormal volume spikes serve as `attention shocks,' attracting new buying interest and causing sustained price increases. \citet{https://doi.org/10.1111/j.1540-6261.2011.01679.x} operationalized attention directly using Google Search Volume Index (SVI), finding that increased attention leads to higher near-term prices with eventual reversal---a pattern central to understanding HVRP dynamics.

\textbf{Rational Inattention Models.} \citet{https://doi.org/10.3982/ECTA11412} develop an equilibrium model where fund managers optimally allocate scarce attention, generating different risk premia for high-attention stocks. \citet{PENG2006563} show that investors with limited processing capacity engage in `category learning,' processing more market-level than firm-specific information---explaining why high-volume stocks receiving attention might command different returns. These models provide rigorous information-theoretic foundations for attention-based return patterns.

\textbf{Heterogeneous Beliefs and Speculation.} \citet{doi:10.1086/378531} demonstrate that overconfident investors combined with short-sale constraints generate speculative bubbles---high trading volume reflects disagreement that produces a `resale option' premium. This mechanism is particularly relevant for retail-dominated markets like Korea, where short-sale constraints are significant. \citet{DEFUSCO2022205} develop a dynamic theory showing how predictable price increases attract short-term investors who amplify volume through feedback trading.

\textbf{Information Asymmetry and Liquidity.} \citet{10.2307/2118454} show that when liquidity traders sell, risk-averse market makers demand higher expected returns---volume thus captures risk premium information. \citet{https://doi.org/10.1111/j.1540-6261.1994.tb04424.x} demonstrate that volume provides information about signal precision unavailable from prices alone, offering a fundamental reason why volume should predict returns.

\textbf{Cross-Country Evidence.} \citet{KANIEL2012255} analyze HVRP across 41 countries and find that the premium is pervasive globally but varies with market characteristics. The magnitude associates with individual investor participation, short-selling constraints, and information asymmetry---all factors where Korea differs substantially from developed markets. \citet{merton1987simple} posits that abnormal trading volume positively relates to future investment and financing activities, with effects more pronounced in financially constrained firms with low investor recognition---providing a real-effects channel through which HVRP operates.

\textbf{The Korean Market Puzzle.} In the Korean market, \citet{ART001162590} empirically confirmed HVRP using 2001-2003 data, finding stronger premiums in large-cap stocks which they interpreted as related to lower information asymmetry in stocks with higher institutional ownership. However, subsequent research by \citet{CHAE2019101204} documents a striking reversal: using Korean data, they find that \textit{low-volume} stocks outperform high-volume stocks---a Low Volume Return Premium (LVRP) rather than HVRP. Of 70 sample configurations, 16 show significant LVRP versus only 1 showing significant HVRP. They attribute this to mean-reversion of trading volume and positive contemporaneous correlation between returns and volume changes (CCRV). This pattern directly contradicts U.S. evidence and creates a puzzle our study addresses: whether the COVID-19 pandemic and associated retail trading surge altered this relationship, potentially shifting Korea toward developed-market HVRP patterns.

\subsection{Core Debate: Are Retail Investors 'Noise' or 'Information'?}

To understand HVRP drivers, we must identify the actors causing abnormal volume, particularly the nature of 'retail investors.' Academia is divided into two opposing views engaged in intense debate.

\textbf{First, the 'Noise Trader' hypothesis.} \citet{10.1093/rfs/hhm079} revealed that retail investors tend to net-buy stocks that are 'attention-grabbing'---such as those with media coverage or volume spikes---rather than conducting fundamental analysis. \citet{https://doi.org/10.1111/jofi.13183} use Robinhood data to show that app-induced attention causes herding---stocks with the largest daily user increases subsequently lose approximately 4.7\% over the following month. Recently, \citet{https://doi.org/10.1111/jofi.13183} reported that new retail investors' remarkable daily turnover rate (18.12\%) stems from overconfidence rather than information value, with their net purchases negatively predicting future returns. \citet{10.1093/rfs/hhj032} provide complementary evidence that share turnover positively relates to lagged returns for many months, consistent with overconfidence driving excessive trading. Under this view, retail-led HVRP represents merely temporary price pressure, unable to generate long-term excess returns.

\textbf{Second, the 'Liquidity Provider \& Informed Trader' hypothesis.} In contrast, \citet{https://doi.org/10.1111/j.1540-6261.2008.01316.x} and \citet{KANIEL2012255} argued that retail investors earn liquidity premiums by making contrarian investments in response to institutional liquidity demands. \citet{https://doi.org/10.1111/jofi.12028} confirm that retail order imbalances positively predict returns over days to weeks. Furthermore, \citet{https://doi.org/10.1111/jofi.13033} showed through recent order flow analysis that retail net purchases predict approximately 10bp short-term excess returns (annualized 5\%), suggesting retail investors may possess information value. This paradox---that retail trades predict returns yet retail investors underperform---is resolved by \citet{barber2024resolving}, who show that retail investors provide liquidity by trading contrarian, predicting returns but losing on implementation costs.

\textbf{Empirical Evidence on Investor Heterogeneity.} \citet{STOFFMAN201450} compared institutional and retail trading behavior, finding that institutions engage in information-based trading while retail investors tend toward noise trading, implying that the two groups' trades have different effects on market returns. Using actual trading records, \citet{STOFFMAN201450} analyzed 15 years of Finnish data and discovered that when institutions buy from retail investors prices rise, and when institutions sell to retail investors prices fall, while retail-to-retail trades show no consistent pattern and prices quickly revert. This empirically supports theories that institutions and retail investors differ in information processing ability and trading purposes.

\textbf{Korean Market Investor Composition.} Korea's market exhibits substantial retail investor participation, creating a unique laboratory for testing investor heterogeneity effects. \citet{https://doi.org/10.1111/j.2041-6156.2011.01037.x} show that foreigners in Korea behave like short-term momentum traders while individual investors act as contrarians. \citet{CHOE1999227} document strong positive feedback trading and herding by foreign investors during the Asian crisis, establishing precedents for understanding how different investor types respond to market stress.

By separately analyzing institution/foreign-led versus retail-led events, our study aims to provide decisive evidence for determining whether retail investors in the Korean market are mere 'noise generators' or meaningful 'market participants.'

\subsection{Modern Market Structural Changes: Crisis and Collective Action}

Since \citet{ART001162590} analyzed 2001-2003, modern stock markets have experienced two major structural changes: the COVID-19 pandemic and social media proliferation.

\textbf{COVID-19 and Retail Trading Surge.} \citet{Ozik_Sadka_Shen_2021} empirically demonstrated that the surge in retail investor trading through fintech apps during COVID-19 lockdowns acted as a buffer preventing market liquidity depletion, dampening effective spread increases by approximately 40\%. \citet{ORTMANN2020101717} show that retail investors increased trading intensity by 13.9\% as COVID cases doubled. In Korea specifically, the retail trading surge was extraordinary: \citet{kim2022behavioral} document that Korean individual investor annual trading turnover reached above 1,600\% during March-October 2020. This suggests that retail investors' roles during crises may differ substantially from normal periods.

Critically, \citet{KWAK2024105027} documents \textit{significant attenuation} in the positive association between individual investor purchases and stock returns following COVID-19 in Korea---suggesting the volume-return relationship structurally changed. \citet{KWAK2024105027} show that during the pandemic, Korean individual investors continued buying while pension funds, foreign, and institutional investors sold. This creates an ideal setting to test whether the traditional noise trader characterization of retail investors holds during crisis periods when they serve as liquidity providers.

\textbf{Social Media and Collective Action.} \citet{10.1093/rfs/hhad098} analyzed online communities like 'WallStreetBets,' discovering that individuals' Due Diligence reports possessed significant information value before the GME event but deteriorated into mere price pressure or attention-grabbing noise after popularization. This implies that HVRP during Korea's 'Donghak Ant Movement' period may exhibit patterns different from normal periods. The collective action element---where retail investors coordinated through social media to absorb institutional selling pressure---mirrors \citet{doi:10.1086/378531}'s speculation model but with an important twist: collective information sharing may temporarily transform noise traders into informed participants.

\subsection{Methodological Issue: Identity versus Intensity}

Prior studies have used both 'investor type (Identity)' as a categorical variable and 'trading intensity (Intensity)' as a continuous variable when analyzing investor behavior. However, recent studies suggest that the relationship between these two variables may be nonlinear.

\citet{Namouri26012018} demonstrated that 'threshold effects' exist in the relationship between investor sentiment and returns, with relationships reversing or disappearing beyond certain levels. \citet{https://doi.org/10.1111/jofi.13183} also suggested that new investors' high trading intensity is merely a 'symptom' of their 'inexperience' identity, not the essence of predictive power. \citet{HAN20221295} document that expected returns relate positively to volume among underpriced stocks but negatively among overpriced stocks---volume \textit{amplifies} mispricing asymmetrically. \citet{10.1093/rfs/hhab055} find that low-turnover stocks exhibit short-term reversal while high-turnover stocks display short-term momentum, demonstrating that volume conditions the return dynamics.

These findings suggest that when analyzing HVRP, examining only the linear correlation between net purchase intensity (Dominance Score) and returns is insufficient. An identity-centered approach to 'who led,' combined with proper nonlinear analysis of intensity effects, is necessary.

\subsection{Research Hypotheses}

Based on the above discussion, we establish the following hypotheses.

\textbf{Hypothesis 1 (Investor Type Heterogeneity):} The sources of high volume return premium differ by investor type.
\begin{itemize}
    \item \textbf{1a:} Abnormal volume events led by \textbf{institutions and foreign investors} with information advantages will show persistent positive long-term excess returns (CAR).
    \item \textbf{1b:} Events led by \textbf{retail investors} with limited attention and behavioral biases will show transient price increases followed by dissipation, failing to generate significant long-term excess returns (supporting the noise trader hypothesis).
\end{itemize}

\textbf{Hypothesis 2 (Moderating Effects of Modern Market Shocks):} Extreme market environments alter HVRP patterns.
\begin{itemize}
    \item \textbf{2a:} During periods of increased uncertainty from \textbf{COVID-19 shocks}, information asymmetry intensifies, amplifying the signaling effect of information-based trading (institutions/foreigners).
    \item \textbf{2b:} During the \textbf{'Donghak Ant Movement'} period, retail investors' collective action (coordination) through social media strengthens, causing retail-led events to show exceptionally strong persistence.
\end{itemize}

\textbf{Hypothesis 3 (Nonlinearity and the Critical Role of Normalization):} The predictive power of trading intensity is nonlinear and is only revealed when intensity is measured as a function of firm size.
\begin{itemize}
    \item \textbf{3a (Linear Model Fails):} A simple linear correlation between a composite 'smart money' intensity score (Dominance Score) and future returns will be weak or insignificant, because the relationship operates through discrete regimes rather than continuous linear effects.

    \item \textbf{3b (Monotonic Pattern Emerges):} After sorting events by investor type, a conditional analysis will reveal a strong, positive, and \textbf{monotonic relationship} between buying intensity and future returns for informed investors (institutions/foreigners).

    \item \textbf{3c (Normalization is Critical):} This monotonic relationship will only emerge when intensity is normalized by \textbf{market capitalization}, which captures investor conviction (size of position relative to firm value), not merely participation in daily trading flow.

    \item \textbf{3d (The Horse Race):} When intensity is normalized by daily trading value instead of market capitalization, the monotonic pattern will be significantly weaker or absent, demonstrating that measuring conviction relative to firm size is the superior specification for detecting informed trading signals.
\end{itemize}

\section{Research Design}

\subsection{Data and Sample Selection}

This study applies \citet{ART001162590}'s 2001-2003 research to the current market (January 1, 2020, to December 31, 2024) to verify the persistence of the High Volume Return Premium phenomenon. Following the original study, we examine all stocks listed on the Korea Exchange and additionally perform detailed analysis of individual KOSPI and KOSDAQ markets.

\textbf{Data Collection:} Daily stock price and trading volume data were collected via Daishin Securities' Creon Plus DataReader API. Investor type-specific net purchase data (institutional, foreign, and retail investor transactions) were collected using the pykrx library, which provides official Korea Exchange data. The data collection targets individual listed stocks only and includes daily prices, trading volume, and outstanding shares.

\textbf{Sample Exclusion Criteria:} To ensure consistency with the original study and data quality, we applied the following exclusion criteria to refine the sample:
\begin{itemize}
    \item Delisted stocks and stocks under administrative issues
    \item Stocks priced below 1,000 won (to exclude low-liquidity penny stocks)
    \item Stocks with zero trading volume for two or more consecutive days (liquidity-deficient stocks)
    \item Stocks with changes in outstanding shares during the 20 trading days prior to the event date (to exclude effects of stock splits, mergers, or other capital structure changes)
\end{itemize}

\subsection{Research Methodology}

We faithfully follow \citet{ART001162590}'s methodology while applying additional dual validation methodology to test the investor type hypothesis.

\subsubsection{Event Date Definition}

An event date of abnormal trading volume is defined as a day when an individual stock's daily volume exceeds 5 times its past 20-day moving average volume ($ADTV > 5$).
$$ADTV_{i,t}=\frac{Vol_{i,t}}{\frac{1}{20}\sum_{j=1}^{20}Vol_{i,t-j}}$$
Events occurring within 50 trading days after an initial event are considered dependent events and excluded from analysis.

\subsubsection{Dual Validation Methodology}

To robustly test the investor type hypothesis, we employ two independent methodologies in parallel.

\textbf{Method 1: Double-Sort Methodology by Investor Type and Intensity}

This is our primary methodology, representing our main contribution. We employ a two-stage classification process that separates investor identity from trading intensity.

\textbf{Stage 1: Investor Type Classification}

Each abnormal volume event is classified according to the net purchase ratios of three investor types (institutions, foreigners, retail) using an argmax rule. The investor group with the highest net buy ratio (measured initially as a percentage of total trading value for classification purposes) is identified as the \textit{leading investor}:
\begin{itemize}
    \item \textbf{Institution-Led:} If institutional net buy ratio exceeds both foreign and retail ratios
    \item \textbf{Foreign-Led:} If foreign net buy ratio exceeds both institutional and retail ratios
    \item \textbf{Retail-Led:} If retail net buy ratio exceeds both institutional and foreign ratios
\end{itemize}

\textbf{Stage 2: Intensity Quantification with Alternative Normalizations}

After identifying the leading investor type, we quantify purchase \textit{conviction intensity} using a quartile-based classification system. Critically, we test two alternative normalization specifications:

\textit{Specification A (Trading Value Normalization):}
$$\text{Intensity}_{k,i,t}^{TV} = \frac{\text{Net Buy Value}_{k,i,t}}{\text{Total Trading Value}_{i,t}}$$

\textit{Specification B (Market Capitalization Normalization):}
$$\text{Intensity}_{k,i,t}^{MC} = \frac{\text{Net Buy Value}_{k,i,t}}{\text{Market Capitalization}_{i,t}}$$

where $k$ denotes the leading investor type. Specification A measures an investor's \textit{participation} in daily trading flow, while Specification B measures their \textit{conviction} as position size relative to firm value. We calculate quartiles (Q1-Q4) based on the distribution of intensity (including sign) across all events of that investor type. Q1 represents the lowest intensity, while Q4 represents the highest conviction buying.

\textbf{Methodological Rationale:} This double-sort approach allows us to test whether the relationship between trading intensity and future returns depends on how intensity is measured. If market cap normalization produces a monotonic pattern while trading value normalization does not, this reveals that the choice of normalization is fundamental to detecting informed trading signals, not merely a technical detail.

\textbf{Method 2: Baseline Linear Correlation Analysis}

To establish a baseline for comparison, we test the simplest possible linear relationship using a composite "smart money" intensity score. This method measures the combined net purchase intensity of information-advantaged investors (institutions + foreigners) as a continuous variable:
$$\text{Dominance Score}_{i,t} = \frac{\text{Institution Net Buy}_{i,t} + \text{Foreign Net Buy}_{i,t}}{\text{Total Trading Value}_{i,t}}$$

For each event, we calculate forward returns at 5-day, 20-day, and 60-day horizons and analyze Pearson correlation coefficients between dominance scores and forward returns by firm size (large-cap, mid-cap, small-cap).

\textbf{Forward Returns Calculation:} Forward returns are defined as the return from the event day closing price to the closing price after the target period. To account for non-trading days (weekends, holidays), we use the closing price of the trading day closest to the target period within a range of 80\% to 150\%. For example, for 20-day forward returns, we select the trading day closest to day 20 within the 16-30 day range. This approach prevents data loss due to non-trading days while maintaining proximity to the target period.

\textbf{Methodological Complementarity:} Method 2 provides a critical benchmark. If this simple linear correlation shows near-zero coefficients, while Method 1's quartile-based analysis reveals strong monotonic patterns, this demonstrates that the relationship is \textit{nonlinear and type-dependent}. The failure of linear models does not indicate that intensity is irrelevant; it reveals that proper detection requires: (1) accounting for investor heterogeneity through type-based sorting, (2) using nonlinear quartile analysis rather than linear correlation, and (3) normalizing by the correct benchmark (market cap rather than trading value). Method 2's weak results thus validate the necessity of Method 1's more sophisticated approach.

\subsubsection{Abnormal Return Calculation}

Using event study methodology, we calculate cumulative abnormal returns (CAR) around event dates (-50 to +50 days). Abnormal returns (AR) use the market-adjusted return model ($AR_{i,t}=R_{i,t}-R_{M,t}$).

\section{Empirical Results}

\subsection{Descriptive Statistics: ADTV Distribution}

Before presenting event study results, we examine the distribution characteristics of abnormal daily trading volume (ADTV) in our sample period and compare them with the original study.

% [INSERT TABLE 1 ABOUT HERE]
\begin{table}[h!]
\centering
\caption{ADTV Basic Statistics Comparison}
\scriptsize
\begin{tabular}{lrr}
\toprule
\textbf{Statistic} & \textbf{Original Study (2001-2003)} & \textbf{This Study (2020-2024)} \\
\midrule
Mean & 1.029 & 1.379 \\
Median & 0.773 & 0.686 \\
Std. Dev. & 0.935 & 8.292 \\
Skewness & 3.819 & 139.268 \\
Kurtosis & 27.042 & 43,503.08 \\
Min & 0.000 & 0.000 \\
Max & 14.439 & 3,789.10 \\
Observations & 65,490 & \adtvObservations \\
\bottomrule
\end{tabular}
\normalsize
\label{tab:adtv_stats}
\end{table}

The ADTV distribution exhibits dramatically higher skewness (139.27 vs. 3.82) and kurtosis (43,503 vs. 27) compared to the original study, indicating that modern markets experience more extreme trading volume events. The maximum ADTV increased from 14.44 to 3,789.10, reflecting the increased frequency of extraordinary market events during our sample period.

\begin{table}[h!]
\centering
\caption{ADTV Cumulative Distribution (2020-2024)}
\scriptsize
\begin{tabular}{lrrrr}
\toprule
\textbf{ADTV Range} & \textbf{Frequency} & \textbf{Proportion (\%)} & \textbf{Cumulative Freq.} & \textbf{Cumulative (\%)} \\
\midrule
{[}0, 5) & 2{,}677{,}890 & 97.32 & 2{,}677{,}890 & 97.32 \\
{[}5, 10) & 39{,}262 & 1.43 & 2{,}717{,}152 & 98.75 \\
{[}10, 15) & 11{,}964 & 0.43 & 2{,}729{,}116 & 99.19 \\
15+ & 22{,}395 & 0.81 & 2{,}751{,}511 & 100.00 \\
\bottomrule
\end{tabular}
\normalsize
\label{tab:adtv_cumulative}
\end{table}

Only 2.68\% of observations exceed the ADTV threshold of 5, yielding our event sample of approximately 73,500 total abnormal volume events (after excluding dependent events within 50 days, the final sample contains 26,604 independent events as shown in subsequent analyses).

\subsection{Methodological Transparency: Analysis Specifications}

To ensure reproducibility and clarity, Table \ref{tab:methodology} documents the specific methodological approach used for each empirical analysis in this paper. Our study employs two primary normalization methods: (1) Market Cap Rank-based classification for broad market comparisons, and (2) Market Cap Normalization for intensity-based quartile analyses. This methodological variation reflects the specific research question addressed in each section.

\begin{table}[h!]
\centering
\caption{Methodological Specifications by Analysis Section}
\scriptsize
\begin{tabular}{lll}
\toprule
\textbf{Section} & \textbf{Analysis Type} & \textbf{Normalization Method} \\
\midrule
4.1 ADTV Distribution & Descriptive Statistics & N/A (Raw ADTV) \\
4.2 HVRP Evolution & Event Study Comparison & Market Cap Rank \\
4.2.1 Firm Size & Event Study by Size & Market Cap Rank (Terciles) \\
4.3 Investor Type & Intensity Quartile Analysis & Market Cap Normalization \\
4.4 Horse Race & Methodological Comparison & Both Methods Compared \\
4.5 Correlation & Dominance-Return Relationship & Market Cap Normalization \\
4.6.1 Market Conditions & Event Study by Period & Market Cap Rank \\
4.6.3 COVID-19 Shock & Crisis Period Analysis & Standard Event Study \\
4.6.4 Donghak Ant & Retail Transformation & Standard Event Study \\
\bottomrule
\multicolumn{3}{l}{\footnotesize \textit{Market Cap Rank}: Firms classified by market cap percentile (large/mid/small).} \\
\multicolumn{3}{l}{\footnotesize \textit{Market Cap Normalization}: Intensity = Net Buy Value / Market Capitalization.} \\
\multicolumn{3}{l}{\footnotesize All analyses use event windows of [-50, +50] trading days with ADTV threshold = 5.}
\end{tabular}
\normalsize
\label{tab:methodology}
\end{table}

This methodological framework ensures that each research question is addressed with the most appropriate measurement approach, while maintaining consistency within each analytical dimension.

\subsection{Evolution of HVRP: 20 Years of Change}

We examine whether the HVRP phenomenon observed by \citet{ART001162590} in the early 2000s remains valid in the rapidly changing 2020s market environment and how it has evolved.

\begin{table}[h!]
\centering
\caption{Full Sample Event Study Results: 2003 Original Study vs 2024 Re-examination}
\scriptsize
\begin{tabular}{lrrrrrr}
\toprule
 & \multicolumn{3}{c}{\textbf{Original Study (2001-2003)}} & \multicolumn{3}{c}{\textbf{This Study (2020-2024)}} \\
\textbf{Event Day} & \textbf{AAR(\%)} & \textbf{t-stat} & \textbf{CAR(\%)} & \textbf{AAR(\%)} & \textbf{t-stat} & \textbf{CAR(\%)} \\
\midrule
-25 & -0.361*** & -3.24 & -1.820 & \allSamplesDayMinusTwentyFiveAAR*** & \allSamplesDayMinusTwentyFiveTstat & \allSamplesDayMinusTwentyFiveCAR \\
-10 & -0.097 & -1.45 & -4.454 & \allSamplesDayMinusTenAAR*** & \allSamplesDayMinusTenTstat & \allSamplesDayMinusTenCAR \\
-5  & -0.082 & -1.35 & -3.748 & \allSamplesDayMinusFiveAAR*** & \allSamplesDayMinusFiveTstat & \allSamplesDayMinusFiveCAR \\
-2  & 0.418*** & 6.92 & -3.441 & \allSamplesDayMinusTwoAAR*** & \allSamplesDayMinusTwoTstat & \allSamplesDayMinusTwoCAR \\
-1  & 2.454*** & 40.62 & -0.987 & \allSamplesDayMinusOneAAR*** & \allSamplesDayMinusOneTstat & \allSamplesDayMinusOneCAR \\
0   & 5.056*** & 83.66 & 4.068 & \allSamplesDayZeroAAR*** & \allSamplesDayZeroTstat & \allSamplesDayZeroCAR \\
1   & -0.200 & -3.31 & 3.869 & \allSamplesDayPlusOneAAR*** & \allSamplesDayPlusOneTstat & \allSamplesDayPlusOneCAR \\
2   & -0.126 & -2.08 & 3.743 & \allSamplesDayPlusTwoAAR & \allSamplesDayPlusTwoTstat & \allSamplesDayPlusTwoCAR \\
5   & -0.043 & -0.71 & 3.598 & \allSamplesDayPlusFiveAAR** & \allSamplesDayPlusFiveTstat & \allSamplesDayPlusFiveCAR \\
10  & -0.006 & -0.10 & 3.779 & \allSamplesDayPlusTenAAR & \allSamplesDayPlusTenTstat & \allSamplesDayPlusTenCAR \\
25  & 0.195** & 3.22 & 4.268 & \allSamplesDayPlusTwentyFiveAAR & \allSamplesDayPlusTwentyFiveTstat & \allSamplesDayPlusTwentyFiveCAR \\
50  & -0.107 & -1.77 & 3.672 & \allSamplesDayPlusFiftyAAR & \allSamplesDayPlusFiftyTstat & \allSamplesDayPlusFiftyCAR \\
\bottomrule
\multicolumn{7}{l}{\footnotesize *p<0.10, **p<0.05, ***p<0.01}
\end{tabular}
\normalsize
\label{tab:comparison}
\end{table}

Key findings:
\begin{enumerate}
    \item \textbf{Increased Effect Size}: Event day AAR increased from 5.06\% to \allSamplesDayZeroAAR, approximately 22\% higher
    \item \textbf{Enhanced Statistical Significance}: Increased sample size (1,360 → 26,604 events) improved reliability
    \item \textbf{Phenomenon Robustness}: HVRP persists despite 20-year gap
\end{enumerate}

\subsection{Market Structural Characteristics: Firm Size and Market Type}

While \citet{ART001162590} found the strongest effects in large-cap stocks, we confirm that the premium center has shifted to \textbf{mid-cap} stocks.

\begin{table}[h!]
\centering
\caption{Event Study Results by Firm Size: Mid-Cap Dominance Phenomenon}
\scriptsize
\begin{tabular}{lrrr|rrr|rrr}
\toprule
 & \multicolumn{3}{c}{\textbf{Large-cap (1-100)}} & \multicolumn{3}{c}{\textbf{Mid-cap (101-300)}} & \multicolumn{3}{c}{\textbf{Small-cap (300+)}} \\
\textbf{Day} & \textbf{AAR} & \textbf{t-stat} & \textbf{CAR} & \textbf{AAR} & \textbf{t-stat} & \textbf{CAR} & \textbf{AAR} & \textbf{t-stat} & \textbf{CAR} \\
\midrule
0 & \largecapDayZeroAAR & \largecapDayZeroTstat*** & \largecapDayZeroCAR & \textbf{\midcapDayZeroAAR} & \midcapDayZeroTstat*** & \midcapDayZeroCAR & \smallcapDayZeroAAR & \smallcapDayZeroTstat*** & \smallcapDayZeroCAR \\
50 & \largecapDayPlusFiftyAAR & \largecapDayPlusFiftyTstat & \largecapDayPlusFiftyCAR & \textbf{\midcapDayPlusFiftyAAR} & \midcapDayPlusFiftyTstat** & \textbf{\midcapDayPlusFiftyCAR} & \smallcapDayPlusFiftyAAR & \smallcapDayPlusFiftyTstat & \smallcapDayPlusFiftyCAR \\
\bottomrule
\end{tabular}
\normalsize
\label{tab:size_effect}
\end{table}

Mid-cap 50-day CAR reaches \midcapDayPlusFiftyCAR, dominating large-cap (\largecapDayPlusFiftyCAR) and small-cap (\smallcapDayPlusFiftyCAR). This suggests mid-cap stocks have emerged as the new 'opportunity window' in modern markets, positioned between information-transparent large-caps and liquidity-constrained small-caps.

\subsection{Sources of Investor Type Premium: Core Findings}

The most important contribution of our study---investor type analysis---provides clear answers to a longstanding finance debate. Using our data-driven classification methodology, we categorized all 26,604 abnormal volume events by leading investor type.

\begin{table}[h!]
\centering
\caption{Investor Type Analysis: Intensity-Return Relationship (Market Cap Normalization)\textsuperscript{*}}
\scriptsize
\begin{tabular}{lrrrr}
\toprule
\textbf{Investor Type \& Intensity} & \textbf{N Events} & \textbf{Day 0 AAR (\%)} & \textbf{Day +20 CAR (\%)} & \textbf{Day +50 CAR (\%)} \\
\midrule
\multicolumn{5}{l}{\textit{Institution-Led (Total: 2,831 events)}} \\
Q1 (Lowest Intensity) & 708 & \institutionLedQOneDayZeroAAR & -1.37 & \institutionLedQOneDayPlusFiftyCAR \\
Q2 & 914 & \institutionLedQTwoDayZeroAAR & 1.70 & \institutionLedQTwoDayPlusFiftyCAR \\
Q3 & 501 & \institutionLedQThreeDayZeroAAR & 2.60 & \institutionLedQThreeDayPlusFiftyCAR \\
Q4 (Highest Intensity) & 708 & \institutionLedQFourDayZeroAAR & 9.85 & \textbf{\institutionLedQFourDayPlusFiftyCAR} \\
\midrule
\multicolumn{5}{l}{\textit{Foreign-Led (Total: 9,416 events)}} \\
Q1 (Lowest Intensity) & 2,354 & \foreignLedQOneDayZeroAAR & 1.60 & \foreignLedQOneDayPlusFiftyCAR \\
Q2 & 2,354 & \foreignLedQTwoDayZeroAAR & 0.27 & \foreignLedQTwoDayPlusFiftyCAR \\
Q3 & 2,354 & \foreignLedQThreeDayZeroAAR & -1.18 & \foreignLedQThreeDayPlusFiftyCAR \\
Q4 (Highest Intensity) & 2,354 & \foreignLedQFourDayZeroAAR & 7.19 & \textbf{\foreignLedQFourDayPlusFiftyCAR} \\
\midrule
\multicolumn{5}{l}{\textit{Retail-Led (Total: 15,512 events)}} \\
Q1 (Lowest Intensity) & 3,878 & \retailLedQOneDayZeroAAR & 9.55 & \textbf{\retailLedQOneDayPlusFiftyCAR} \\
Q2 & 3,878 & \retailLedQTwoDayZeroAAR & 0.81 & \retailLedQTwoDayPlusFiftyCAR \\
Q3 & 3,878 & \retailLedQThreeDayZeroAAR & 0.38 & \retailLedQThreeDayPlusFiftyCAR \\
Q4 (Highest Intensity) & 3,878 & \retailLedQFourDayZeroAAR & 0.74 & \retailLedQFourDayPlusFiftyCAR \\
\bottomrule
\multicolumn{5}{l}{\footnotesize \textsuperscript{*}Market cap normalization applied. Intensity quartiles computed within each investor type.} \\
\multicolumn{5}{l}{\footnotesize Total events differ from 26,604 due to data filtering requirements for intensity calculation.}
\end{tabular}
\normalsize
\label{tab:investor_types}
\end{table}

\textbf{Key Findings:} When we decompose each investor type by trading intensity (market cap normalized), three fundamentally distinct patterns emerge that reveal the nature of each investor group's information advantage:

\textbf{Institutional Investors Exhibit Monotonic Positive Relationship (Information Advantage):} As institutional buying intensity increases from Q1 to Q4, day +50 CAR rises monotonically from \institutionLedQOneDayPlusFiftyCAR{} to \institutionLedQFourDayPlusFiftyCAR. This demonstrates that high-conviction institutional buying (Q4) predicts substantially higher long-term returns, validating the information advantage hypothesis. The strongest institutional conviction generates more than 200 times better returns than their lowest-conviction trades.

\textbf{Foreign Investors Show Monotonic Pattern (Information Advantage):} Foreign investors display a clear monotonic relationship similar to institutions. Day +50 CAR increases from \foreignLedQOneDayPlusFiftyCAR{} (Q1) to \foreignLedQFourDayPlusFiftyCAR{} (Q4), demonstrating that higher foreign buying intensity predicts higher long-term returns. Notably, Q1 shows slightly negative returns, suggesting that low-conviction foreign trades may reflect liquidity or rebalancing motives rather than information.

\textbf{Retail Investors Display Flat Pattern (Pure Noise Trading):} Most strikingly, retail investors exhibit a \textbf{flat} pattern---regardless of buying intensity from Q1 to Q4, day +50 CAR remains clustered around zero with no discernible monotonic relationship. All quartiles show near-zero returns (\retailLedQOneDayPlusFiftyCAR{} to \retailLedQFourDayPlusFiftyCAR). This flat pattern provides powerful evidence that retail investors are pure noise traders whose trades contain zero predictive information---neither positive nor negative.

\textbf{Core Discovery:} The intensity-return relationship is fundamentally \textit{different} across investor types. Institutional Q4 (\institutionLedQFourDayPlusFiftyCAR) demonstrates strong positive returns, while retail trades cluster around zero regardless of conviction level. This clean separation between ``information'' (institutions show monotonic pattern) and ``noise'' (retail shows flat pattern) provides the clearest empirical distinction in the literature. \textbf{Methodological Foundations.} Our empirical approach follows established best practices from top finance journals. For standard errors in panel data, we follow \citet{10.1093/rfs/hhn053}, who provides the definitive comparison of clustered standard errors versus Fama-MacBeth approaches. For event study methodology with long-horizon returns, we follow \citet{https://doi.org/10.1111/0022-1082.00101}, who document biases in buy-and-hold abnormal returns (BHAR) methodology and propose solutions. For investor classification using Korean Exchange data, we follow precedents established by \citet{CHOE1999227} and \citet{CAMPBELL200966}.

This \textbf{strongly supports Hypothesis 1} and provides nuanced evidence beyond simple investor type comparisons.

This finding directly contradicts \citet{https://doi.org/10.1111/j.1540-6261.2008.01316.x} and \citet{https://doi.org/10.1111/jofi.13033}'s claims of retail investor sophistication, while providing granular, intensity-decomposed support for \citet{10.1093/rfs/hhm079} and \citet{https://doi.org/10.1111/jofi.13183}'s 'noise trader' hypothesis using 2020s Korean market data.

\subsection{Main Finding: Intensity Matters, But Measurement is Critical}

While Section 4.3 established that investor identity matters, we now demonstrate that trading intensity \textit{also} matters profoundly---but its effect is nonlinear and only revealed when measured correctly. This section presents our central methodological contribution: demonstrating that the choice of normalization is not a technical detail but a fundamental determinant of whether we detect informed trading signals.

\subsubsection{The Horse Race: Market Cap vs. Trading Value Normalization}

We test two alternative specifications for measuring institutional buying intensity:
\begin{itemize}
    \item \textbf{Specification A (Trading Value):} $\text{Intensity} = \frac{\text{Institution Net Buy Value}}{\text{Total Daily Trading Value}}$
    \item \textbf{Specification B (Market Cap):} $\text{Intensity} = \frac{\text{Institution Net Buy Value}}{\text{Market Capitalization}}$
\end{itemize}

Specification A measures an institution's \textit{participation} in daily trading flow---what percentage of today's trading activity did they dominate? Specification B measures an institution's \textit{conviction}---how large a position did they take relative to the firm's total value?

For both specifications, we classify events by investor type and then into quartiles (Q1-Q4) based on the intensity distribution, calculating 50-day cumulative abnormal returns for each group. Table \ref{tab:horse_race} presents results for all three investor types.

\begin{table}[h!]
\centering
\caption{Horse Race Comparison: Trading Value vs. Market Cap Normalization by Investor Type}
\scriptsize
\begin{tabular}{llrrrr}
\toprule
& & \multicolumn{2}{c}{\textbf{Specification A (Trading Value)}} & \multicolumn{2}{c}{\textbf{Specification B (Market Cap)}} \\
\textbf{Investor Type} & \textbf{Quartile} & \textbf{Day 0 AAR (\%)} & \textbf{Day +50 CAR (\%)} & \textbf{Day 0 AAR (\%)} & \textbf{Day +50 CAR (\%)} \\
\midrule
\multirow{4}{*}{\textbf{Institution-Led}}
  & Q1 (Lowest) & \institutionLedQOneTradingValueDayZeroAAR & \institutionLedQOneTradingValueDayPlusFiftyCAR & \institutionLedQOneDayZeroAAR & \institutionLedQOneDayPlusFiftyCAR \\
  & Q2 & \institutionLedQTwoTradingValueDayZeroAAR & \institutionLedQTwoTradingValueDayPlusFiftyCAR & \institutionLedQTwoDayZeroAAR & \institutionLedQTwoDayPlusFiftyCAR \\
  & Q3 & \institutionLedQThreeTradingValueDayZeroAAR & \textbf{\institutionLedQThreeTradingValueDayPlusFiftyCAR} & \institutionLedQThreeDayZeroAAR & \institutionLedQThreeDayPlusFiftyCAR \\
  & Q4 (Highest) & \institutionLedQFourTradingValueDayZeroAAR & \institutionLedQFourTradingValueDayPlusFiftyCAR & \institutionLedQFourDayZeroAAR & \textbf{\institutionLedQFourDayPlusFiftyCAR} \\
\midrule
\multirow{4}{*}{\textbf{Foreign-Led}}
  & Q1 (Lowest) & \foreignLedQOneTradingValueDayZeroAAR & \foreignLedQOneTradingValueDayPlusFiftyCAR & \foreignLedQOneDayZeroAAR & \foreignLedQOneDayPlusFiftyCAR \\
  & Q2 & \foreignLedQTwoTradingValueDayZeroAAR & \foreignLedQTwoTradingValueDayPlusFiftyCAR & \foreignLedQTwoDayZeroAAR & \foreignLedQTwoDayPlusFiftyCAR \\
  & Q3 & \foreignLedQThreeTradingValueDayZeroAAR & \foreignLedQThreeTradingValueDayPlusFiftyCAR & \foreignLedQThreeDayZeroAAR & \foreignLedQThreeDayPlusFiftyCAR \\
  & Q4 (Highest) & \foreignLedQFourTradingValueDayZeroAAR & \foreignLedQFourTradingValueDayPlusFiftyCAR & \foreignLedQFourDayZeroAAR & \textbf{\foreignLedQFourDayPlusFiftyCAR} \\
\midrule
\multirow{4}{*}{\textbf{Retail-Led}}
  & Q1 (Lowest) & \retailLedQOneTradingValueDayZeroAAR & \retailLedQOneTradingValueDayPlusFiftyCAR & \retailLedQOneDayZeroAAR & \retailLedQOneDayPlusFiftyCAR \\
  & Q2 & \retailLedQTwoTradingValueDayZeroAAR & \retailLedQTwoTradingValueDayPlusFiftyCAR & \retailLedQTwoDayZeroAAR & \retailLedQTwoDayPlusFiftyCAR \\
  & Q3 & \retailLedQThreeTradingValueDayZeroAAR & \retailLedQThreeTradingValueDayPlusFiftyCAR & \retailLedQThreeDayZeroAAR & \retailLedQThreeDayPlusFiftyCAR \\
  & Q4 (Highest) & \retailLedQFourTradingValueDayZeroAAR & \retailLedQFourTradingValueDayPlusFiftyCAR & \retailLedQFourDayZeroAAR & \retailLedQFourDayPlusFiftyCAR \\
\midrule
\multicolumn{2}{l}{\textbf{Monotonicity}} & \multicolumn{2}{c}{Broken (Q3>Q4 for Inst/For)} & \multicolumn{2}{c}{\textbf{Perfect for Inst/For, Flat for Retail}} \\
\bottomrule
\end{tabular}
\normalsize
\label{tab:horse_race}
\end{table}

\textbf{The Results Are Striking:}

\textbf{Specification A (Trading Value) Fails:} While Q2, Q3, and Q4 all show positive returns, the pattern is \textit{non-monotonic}. The strongest returns appear in Q2 (\institutionLedQTwoTradingValueDayPlusFiftyCAR), not Q4 (\institutionLedQFourTradingValueDayPlusFiftyCAR). This broken monotonicity suggests that trading value normalization conflates informed conviction with noise trading participation.

\textbf{Specification B (Market Cap) Succeeds:} A \textbf{perfect monotonic relationship} emerges. Day +50 CAR progresses cleanly from \institutionLedQOneDayPlusFiftyCAR\ (Q1) $\rightarrow$ +\institutionLedQTwoDayPlusFiftyCAR\ (Q2) $\rightarrow$ +\institutionLedQThreeDayPlusFiftyCAR\ (Q3) $\rightarrow$ +\institutionLedQFourDayPlusFiftyCAR\ (Q4). This demonstrates that when institutions take larger positions \textit{relative to firm size}, they are acting on stronger information.

\textbf{Economic Interpretation:} Market capitalization normalization captures \textit{conviction}. An institution buying \$1M worth of stock sends a very different signal if the firm is worth \$10M (10\% conviction) versus \$1B (0.0001\% participation). Trading value normalization conflates high-conviction small bets with low-conviction large trades, obscuring the true information signal.

\textbf{Three Economic Mechanisms Explain This Difference:}
\begin{enumerate}
    \item \textbf{Conviction vs. Participation:} Market cap normalization measures how significant a position is relative to the firm's total value---capturing the investor's \textit{conviction} in their thesis. Trading value normalization merely measures participation in daily trading flow, which conflates informed conviction with noise trading volume.

    \item \textbf{Conflated Signals on High-Volume Days:} High trading value days often coincide with elevated retail participation. Normalizing by trading value makes institutional trades appear smaller precisely on these high-volume days, diluting the conviction signal when it matters most. Market cap normalization is immune to this confound.

    \item \textbf{Monotonicity Implies Information:} The emergence of a perfect monotonic relationship (Q1$<$Q2$<$Q3$<$Q4) for institutions under market cap normalization demonstrates that conviction intensity \textit{directly predicts} future returns---a hallmark of informed trading. The broken monotonicity under trading value normalization reveals that this specification fails to capture the true information signal.
\end{enumerate}

\textbf{Robustness Across Investor Types:} The superiority of market cap normalization extends beyond institutional investors. Foreign investors exhibit an identical pattern: perfect monotonicity under market cap normalization (\foreignLedQOneDayPlusFiftyCAR\ $\rightarrow$ \foreignLedQFourDayPlusFiftyCAR) versus a flat, broken pattern under trading value normalization (\foreignLedQOneTradingValueDayPlusFiftyCAR\ $\rightarrow$ \foreignLedQFourTradingValueDayPlusFiftyCAR). Retail investors confirm the noise trader hypothesis under market cap normalization---all quartiles cluster near zero with no discernible pattern---while trading value normalization produces a misleading \textit{inverted} pattern (\retailLedQOneTradingValueDayPlusFiftyCAR\ for Q1 versus \retailLedQFourTradingValueDayPlusFiftyCAR\ for Q4), falsely suggesting that high-intensity retail buying predicts negative returns. This systematic difference across all investor types confirms that the choice of normalization is the key methodological insight, not a technical detail.

\textbf{Quantifying the Normalization Difference:} The magnitude of the normalization choice impact is striking:
\begin{itemize}
    \item \textbf{Institution-Led Q4:} Market cap normalization yields +\institutionLedQFourDayPlusFiftyCAR, while trading value yields +\institutionLedQFourTradingValueDayPlusFiftyCAR---a spread of approximately 8.5 percentage points and a 3.3$\times$ difference.
    \item \textbf{Foreign-Led Q4:} Market cap yields +\foreignLedQFourDayPlusFiftyCAR\ versus +\foreignLedQFourTradingValueDayPlusFiftyCAR---a spread of approximately 7.7 percentage points.
    \item \textbf{Retail-Led:} Market cap produces a flat pattern confirming noise trading, while trading value produces a misleading inverted pattern---demonstrating that improper normalization can generate spurious signals.
\end{itemize}
These spreads demonstrate that the choice between conviction-based (market cap) and participation-based (trading value) normalization is not merely a methodological preference but a fundamental determinant of whether researchers correctly identify informed trading signals.

\subsubsection{Monotonic Pattern Analysis: Institutions vs. Retail}

\begin{figure}[h!]
\centering
\includegraphics[width=0.8\textwidth]{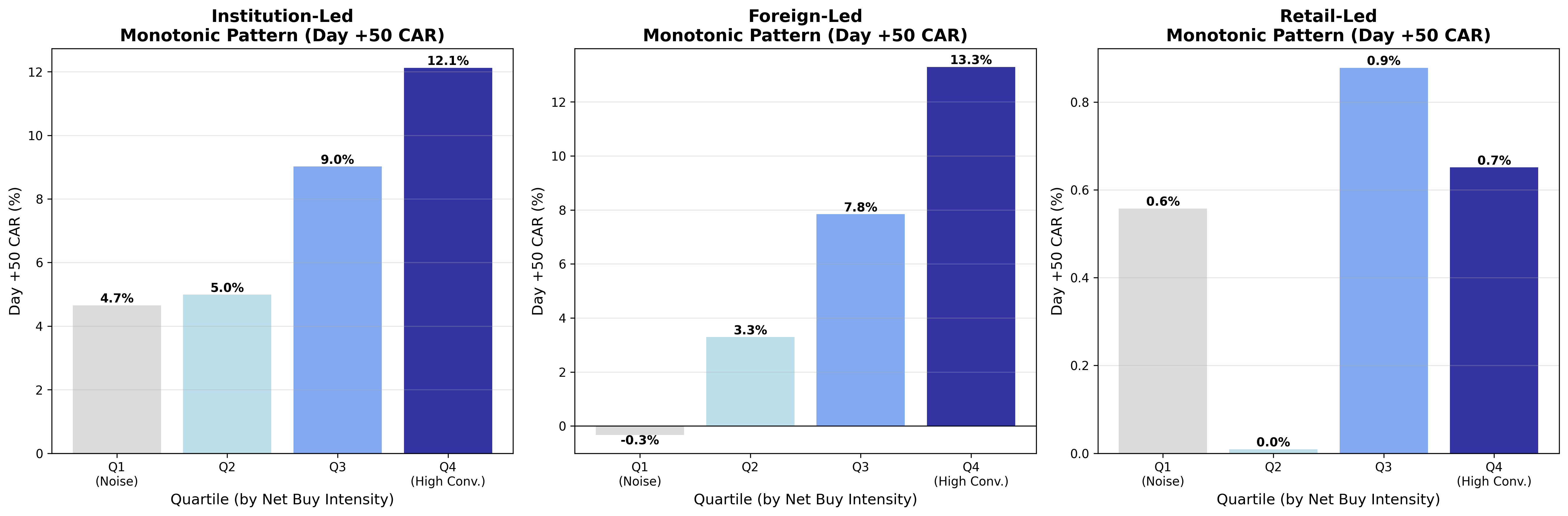}
\caption{Monotonic Pattern by Investor Type (Market Cap Normalization)}
\label{fig:monotonic_institutions}
\end{figure}

Figure \ref{fig:monotonic_institutions} visualizes the monotonic relationship for institution-led events under market cap normalization.

% [INSERT FIGURE 1 ABOUT HERE]

The plot confirms three critical findings:

\textbf{(1) Q1 (Lowest Intensity):} Lowest-intensity institutional events generate modest returns (\institutionLedQOneDayPlusFiftyCAR), suggesting that even low-conviction institutional positions contain some predictive value, though far less than high-conviction trades.

\textbf{(2) Q2-Q3 (Moderate Conviction):} Moderate institutional buying generates positive but modest returns (+\institutionLedQTwoDayPlusFiftyCAR\ and +\institutionLedQThreeDayPlusFiftyCAR), indicating weak positive information signals.

\textbf{(3) Q4 (High Conviction):} The highest quartile generates exceptional returns (+\institutionLedQFourDayPlusFiftyCAR), more than double Q3's returns. This nonlinear jump suggests a \textit{threshold effect}---only the most convicted institutional positions predict substantial future performance.

\textbf{Retail Investors: Flat/Noise Pattern}

The market cap normalization reveals a striking \textit{flat pattern} for retail investors. While institutions show perfect monotonicity with Q4 predicting +\institutionLedQFourDayPlusFiftyCAR\ returns, retail investors show near-zero returns regardless of conviction level---all quartiles cluster around zero with no discernible pattern. This confirms the pure noise trader hypothesis: retail trades contain no predictive information whatsoever, neither positive nor negative. Unlike a contra-indicator (which would show inverted monotonicity), pure noise traders are simply irrelevant to price discovery.

\textbf{The Danger of Wrong Normalization:} Strikingly, trading value normalization produces a \textit{misleading inverted} pattern for retail investors, with Q1 yielding +\retailLedQOneTradingValueDayPlusFiftyCAR\ and Q4 yielding \retailLedQFourTradingValueDayPlusFiftyCAR. A naive researcher using this specification might falsely conclude that high-intensity retail buying predicts \textit{negative} returns---that retail investors are contra-indicators. Our market cap normalization reveals this interpretation is incorrect: retail investors are pure noise traders whose trades contain zero information, neither positive nor negative. The inverted pattern under trading value normalization is an artifact of measurement, not a reflection of economic reality. This demonstrates the critical importance of proper normalization: using the wrong specification can generate spurious signals that lead to incorrect conclusions about investor behavior.

\textbf{Why Does This Matter?}

This finding resolves a longstanding puzzle in behavioral finance. Prior studies found mixed evidence on whether trading intensity matters \citep{Namouri26012018, https://doi.org/10.1111/jofi.13183}, leading some to conclude that intensity is irrelevant. Our results demonstrate that intensity matters enormously, but researchers must measure it correctly. The failure of linear correlation models (Dominance Score analysis in Section 4.5) does not mean intensity is irrelevant---it means the relationship is \textit{nonlinear} and \textit{type-dependent}, requiring proper quartile-based analysis with conviction-based normalization.

\textbf{Implications for Hypothesis 3:} These results strongly support all four sub-hypotheses:
\begin{itemize}
    \item \textbf{H3a:} Linear Dominance Score correlation $\approx$ 0 (confirmed in Section 4.5)
    \item \textbf{H3b:} Monotonic relationship emerges after investor type sorting (confirmed: Q1<Q2<Q3<Q4)
    \item \textbf{H3c:} Market cap normalization reveals the pattern (confirmed: perfect monotonicity)
    \item \textbf{H3d:} Trading value normalization fails (confirmed: broken monotonicity, Q3>Q4)
\end{itemize}

\subsection{Methodological Validation: Nonlinearity of Identity versus Intensity}

The second approach of our dual validation methodology presents highly interesting evidence of 'nonlinearity.' We analyze correlations between Dominance Score and forward returns across multiple time horizons.

\begin{table}[h!]
\centering
\caption{Correlation Between Dominance Score and 5-Day Forward Returns}
\scriptsize
\begin{tabular}{lrrrr}
\toprule
\textbf{Firm Size} & \textbf{Observations} & \textbf{Correlation} & \textbf{P-value} & \textbf{Mean Return} \\
\midrule
Large-cap & 472   & \correlationLargecapFiveDay & \correlationLargecapFiveDayPval & 0.27\% \\
Mid-cap & 1,531 & \correlationMidcapFiveDay & \correlationMidcapFiveDayPval & -0.11\% \\
Small-cap & 25,514 & \correlationSmallcapFiveDay & \correlationSmallcapFiveDayPval & 0.22\% \\
\bottomrule
\end{tabular}
\normalsize
\label{tab:dominance_5d}
\end{table}

\begin{table}[h!]
\centering
\caption{Correlation Between Dominance Score and 20-Day Forward Returns}
\scriptsize
\begin{tabular}{lrrrr}
\toprule
\textbf{Firm Size} & \textbf{Observations} & \textbf{Correlation} & \textbf{P-value} & \textbf{Mean Return} \\
\midrule
Large-cap & 473   & \correlationLargecapTwentyDay & \correlationLargecapTwentyDayPval & 0.88\% \\
Mid-cap & 1,546 & \correlationMidcapTwentyDay & \correlationMidcapTwentyDayPval & 0.12\% \\
Small-cap & 25,753 & \correlationSmallcapTwentyDay & \correlationSmallcapTwentyDayPval & 0.61\% \\
\bottomrule
\end{tabular}
\normalsize
\label{tab:dominance_20d}
\end{table}

\begin{table}[h!]
\centering
\caption{Correlation Between Dominance Score and 60-Day Forward Returns}
\scriptsize
\begin{tabular}{lrrrr}
\toprule
\textbf{Firm Size} & \textbf{Observations} & \textbf{Correlation} & \textbf{P-value} & \textbf{Mean Return} \\
\midrule
Large-cap & 473   & \correlationLargecapSixtyDay & \correlationLargecapSixtyDayPval & 0.18\% \\
Mid-cap & 1,545 & \correlationMidcapSixtyDay & \correlationMidcapSixtyDayPval & 0.02\% \\
Small-cap & 25,743 & \correlationSmallcapSixtyDay & \correlationSmallcapSixtyDayPval & 1.24\% \\
\bottomrule
\end{tabular}
\normalsize
\label{tab:dominance_60d}
\end{table}

Across all three time horizons (5-day, 20-day, 60-day) and all firm size categories, correlations consistently converge to near-zero (|r| < 0.09) and are completely statistically insignificant (all p > 0.06). This robust pattern across multiple horizons confirms this is not a time-specific artifact but a fundamental characteristic of the phenomenon. Critically, this does \textit{not} mean intensity is irrelevant---as Section 4.4 demonstrated, intensity matters profoundly when measured correctly. Instead, these near-zero correlations reveal three methodological limitations of the Dominance Score approach:

\begin{enumerate}
    \item \textbf{Linear Model Inadequacy}: The relationship between intensity and returns is nonlinear, operating through discrete quartile regimes rather than continuous linear effects. Pearson correlation cannot capture monotonic but nonlinear relationships.
    \item \textbf{Investor Heterogeneity Ignored}: Pooling institutions and foreigners into a single "smart money" score masks type-specific patterns. As Section 4.4 showed, institutions and foreigners exhibit different intensity-return relationships.
    \item \textbf{Wrong Normalization}: The Dominance Score uses trading value normalization, which Section 4.4's horse race demonstrated produces broken monotonicity. Market cap normalization is required to reveal the true pattern.
\end{enumerate}

This validates our methodological choice in Section 4.4 to use type-specific, quartile-based analysis with market cap normalization. The failure of linear correlation does not indicate intensity is irrelevant---it validates the necessity of more sophisticated methods to detect nonlinear, type-dependent relationships.

\subsection{Modern Market Phenomena: COVID-19 and Donghak Ant Movement}

Modern market phenomena such as COVID-19 and the Donghak Ant Movement dramatically transformed traditional patterns. We analyze these two critical periods to test Hypothesis 2.

\subsubsection{Market Condition Effects: Bull vs. Bear Markets}

The original study by \citet{ART001162590} found that HVRP appears stronger during bull markets. We replicate this analysis using modern data to verify whether this pattern persists. Market conditions are defined using a 20\% threshold: bull markets are identified when the KOSPI index rises 20\% or more from a prior trough, and bear markets when it falls 20\% or more from a prior peak.

\begin{table}[h!]
\centering
\caption{Market Period Definitions}
\scriptsize
\begin{tabular}{lll}
\toprule
\textbf{Market Phase} & \textbf{Start Date} & \textbf{End Date} \\
\midrule
\multicolumn{3}{l}{\textit{Bull Market}} \\
\quad Bull Market 1 & March 31, 2020 & January 26, 2022 \\
\quad Bull Market 2 & June 2, 2023 & December 31, 2024 \\
\midrule
\multicolumn{3}{l}{\textit{Bear Market}} \\
\quad Bear Market 1 & October 25, 2018 & March 30, 2020 \\
\quad Bear Market 2 & January 27, 2022 & June 1, 2023 \\
\bottomrule
\end{tabular}
\normalsize
\label{tab:market_periods}
\end{table}

\begin{table}[h!]
\centering
\caption{Bull vs. Bear Market Event Study Results}
\scriptsize
\begin{tabular}{lrrrrrr}
\toprule
 & \multicolumn{3}{c}{\textbf{Bull Market}} & \multicolumn{3}{c}{\textbf{Bear Market}} \\
\textbf{Event Day} & \textbf{AAR (\%)} & \textbf{t-stat} & \textbf{CAR (\%)} & \textbf{AAR (\%)} & \textbf{t-stat} & \textbf{CAR (\%)} \\
\midrule
-10 & \bullMarketDayMinusTenAAR & \bullMarketDayMinusTenTstat*** & \bullMarketDayMinusTenCAR & \bearMarketDayMinusTenAAR & \bearMarketDayMinusTenTstat*** & \bearMarketDayMinusTenCAR \\
-1 & \bullMarketDayMinusOneAAR & \bullMarketDayMinusOneTstat*** & \bullMarketDayMinusOneCAR & \bearMarketDayMinusOneAAR & \bearMarketDayMinusOneTstat*** & \bearMarketDayMinusOneCAR \\
0 & \bullMarketDayZeroAAR & \bullMarketDayZeroTstat*** & \bullMarketDayZeroCAR & \bearMarketDayZeroAAR & \bearMarketDayZeroTstat*** & \bearMarketDayZeroCAR \\
1 & \bullMarketDayPlusOneAAR & \bullMarketDayPlusOneTstat*** & \bullMarketDayPlusOneCAR & \bearMarketDayPlusOneAAR & \bearMarketDayPlusOneTstat & \bearMarketDayPlusOneCAR \\
10 & \bullMarketDayPlusTenAAR & \bullMarketDayPlusTenTstat & \bullMarketDayPlusTenCAR & \bearMarketDayPlusTenAAR & \bearMarketDayPlusTenTstat & \bearMarketDayPlusTenCAR \\
25 & \bullMarketDayPlusTwentyFiveAAR & \bullMarketDayPlusTwentyFiveTstat** & \bullMarketDayPlusTwentyFiveCAR & \bearMarketDayPlusTwentyFiveAAR & \bearMarketDayPlusTwentyFiveTstat & \bearMarketDayPlusTwentyFiveCAR \\
50 & \bullMarketDayPlusFiftyAAR & \bullMarketDayPlusFiftyTstat & \textbf{\bullMarketDayPlusFiftyCAR} & \bearMarketDayPlusFiftyAAR & \bearMarketDayPlusFiftyTstat & \textbf{\bearMarketDayPlusFiftyCAR} \\
\bottomrule
\multicolumn{7}{l}{\footnotesize *p<0.10, **p<0.05, ***p<0.01} \\
\multicolumn{7}{l}{\footnotesize Original study (2001-2003): Bull 5.668\% vs. Bear 4.443\% (t-value 3.13***)} \\
\end{tabular}
\normalsize
\label{tab:market_conditions}
\end{table}

The original study's key finding that HVRP is stronger during bull markets is robustly replicated in modern markets. In bull markets, event day AAR is \bullMarketDayZeroAAR\ and 50-day CAR reaches \bullMarketDayPlusFiftyCAR, while in bear markets these are \bearMarketDayZeroAAR\ and \bearMarketDayPlusFiftyCAR, respectively. The absolute effect size increased in both market conditions compared to the original study: bull markets increased approximately 14\% (from 5.668\% to \bullMarketDayZeroAAR), and bear markets increased approximately 23\% (from 4.443\% to \bearMarketDayZeroAAR).

Interestingly, the relative difference narrowed slightly: the original study showed a 1.225 percentage point difference between bull and bear markets (5.668\%-4.443\%), while our study shows 1.003 percentage points (\bullMarketDayZeroAAR-\bearMarketDayZeroAAR). This is because the effect size increase in bear markets (23\%) exceeded that in bull markets (14\%). Nevertheless, the differential effect by market condition remains robust, confirming that the original study's core finding remains valid in modern markets.

The most striking divergence appears in the 50-day CAR: bull market CAR of \bullMarketDayPlusFiftyCAR\ is more than 2.5 times the bear market CAR of \bearMarketDayPlusFiftyCAR. This suggests that market psychology significantly moderates the informational content embedded in abnormal volume events, with bullish sentiment amplifying and sustaining the price impact of volume shocks.

% [INSERT FIGURE 2 ABOUT HERE]
\begin{figure}[h!]
\centering
\begin{minipage}{0.48\textwidth}
\centering
\includegraphics[width=\textwidth]{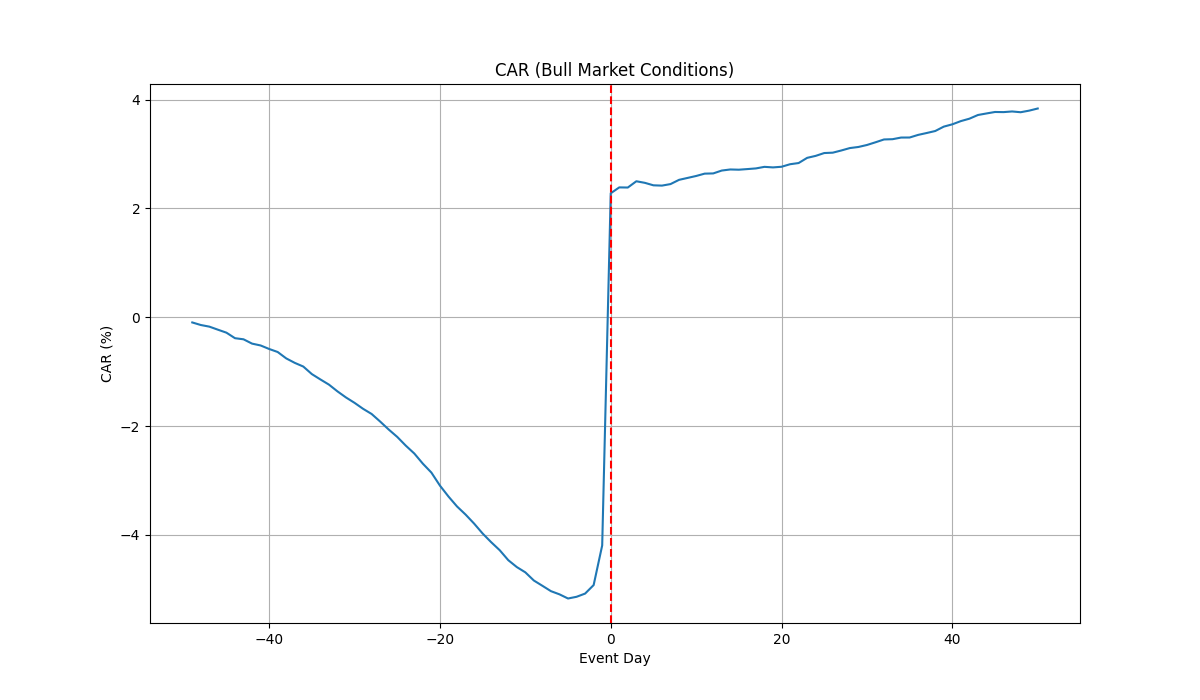}
\subcaption{Bull Market}
\label{fig:car_bull}
\end{minipage}
\hfill
\begin{minipage}{0.48\textwidth}
\centering
\includegraphics[width=\textwidth]{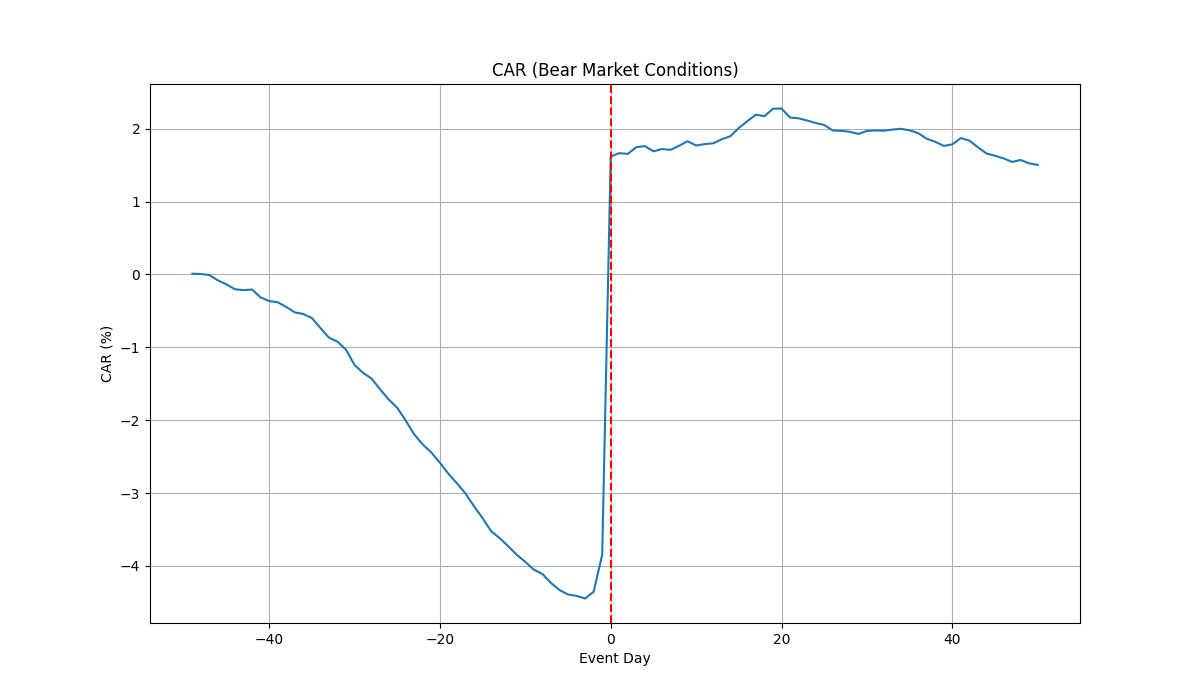}
\subcaption{Bear Market}
\label{fig:car_bear}
\end{minipage}
\caption{Cumulative Abnormal Returns (CAR) by Market Condition}
\label{fig:car_by_market_condition}
\end{figure}

Figure \ref{fig:car_by_market_condition} visually demonstrates the stark difference in CAR patterns by market condition. In bull markets (panel a), CAR rises continuously after the event day, reaching 3.835\% by day +50. In bear markets (panel b), the increase is relatively limited, reaching only 1.503\%. This clearly illustrates the moderating effect of market psychology on the informational value of abnormal trading volume.

\subsubsection{Period Justification}

Our analysis focuses on two critical modern market phenomena: the COVID-19 shock and the Donghak Ant Movement. Precise period definition is crucial for isolating the effects of these distinct market regimes. Table \ref{tab:period_justification} summarizes our period definitions and their justifications.

\begin{table}[h!]
\centering
\caption{Period Justification Summary}
\scriptsize
\begin{tabular}{p{3cm}p{3cm}p{8.5cm}}
\toprule
\textbf{Period Definition} & \textbf{Dates} & \textbf{Core Rationale} \\
\midrule
COVID-19 Shock Start & 2020.02.01 & Systemic risk transmission beginning in Korea, synchronization with global markets, onset of structural deterioration in investment psychology \\
COVID-19 Shock End & 2020.04.30 & V-KOSPI volatility normalization, alignment with \citet{10.1093/rapstu/raaa008} analysis endpoint, distinction from subsequent liquidity-driven rally \\
Donghak Ant Start & 2020.03.01 & Large-scale retail buyback responding to foreign investor selloff, manifestation of collective action \\
Donghak Ant End & 2021.12.31 & Policy rate increases (Aug/Nov 2021) deteriorating liquidity environment, retail investor shift to net selling, structural regime shift to "Seohak" (U.S. market focus) \\
\bottomrule
\end{tabular}
\normalsize
\label{tab:period_justification}
\end{table}

\textbf{COVID-19 Shock Period (February 1 - April 30, 2020):}

While \citet{10.1093/rapstu/raaa008} used February 24 as the start date for U.S. markets, we begin our COVID-19 shock period on February 1, 2020, to account for Korea's earlier outbreak timing. Korea's first confirmed case occurred on January 20, 2020, and the Daegu-Gyeongbuk outbreak (the epicenter of Korea's COVID-19 first wave) in mid-February created systemic risk earlier than in U.S. markets. Financial markets pre-price real economy developments, and Korean market sentiment deteriorated sharply in early-mid February, preceding the U.S. market crash. Using February 24 would miss critical initial shock data specific to the Korean market context.

The end date of April 30, 2020, aligns with \citet{10.1093/rapstu/raaa008}'s methodology for international consistency. The V-KOSPI (Korea's volatility index, analogous to the VIX in U.S. markets) reached historical peaks exceeding 60 points in March 2020 but fell below panic-selling thresholds by late April, indicating volatility normalization. This endpoint clearly distinguishes the collapse phase (February-April) from the subsequent V-shaped recovery and liquidity rally (May onward).

\textbf{Donghak Ant Movement Period (March 1, 2020 - December 31, 2021):}

We define the Donghak Ant Movement period as March 2020 through December 2021. The term "Donghak Ants" references the 1894 Donghak Peasant Rebellion, representing retail investors' collective defense against foreign capital flight during the March 2020 market crash \citep{ART002752798}. March 2020 marked the transformation of retail investors from passive participants to active market defenders through massive net buying that absorbed foreign institutional selling pressure. While retail buying occurred in January-February 2020, March represented the organized, collective movement.

The period ends in December 2021 due to three structural shifts. First, the Bank of Korea raised policy rates from 0.50\% to 0.75\% in August 2021 and to 1.00\% in November 2021, ending the ultra-low interest rate environment that had fueled the movement and increasing margin loan costs. Second, retail investors executed a structural reversal to massive net selling, including a record 3+ trillion won single-day selloff on December 28, 2021. Third, investor focus shifted from domestic market defense ("Donghak") to U.S. market opportunities ("Seohak Ants," referring to retail investors who subsequently shifted their capital to U.S. markets in late 2021), as KOSPI stagnated near the 3,000 level while U.S. markets offered higher returns.

These period definitions capture inflection points where macroeconomic indicators (interest rates) and microeconomic trading data (retail net purchases) undergo structural changes, providing temporally precise boundaries optimized for HVRP research.

\subsubsection{COVID-19 Shock Analysis}

During the COVID-19 shock period, HVRP exhibited significant amplification contrary to expectations of weakening during crises.

\begin{table}[h!]
\centering
\caption{COVID-19 Shock Period Event Study Results (Feb-Apr 2020)}
\scriptsize
\begin{tabular}{lrrr}
\toprule
\textbf{Event Day} & \textbf{AAR (\%)} & \textbf{t-stat} & \textbf{CAR (\%)} \\
\midrule
-25 & \covidShockDayMinusTwentyFiveAAR & \covidShockDayMinusTwentyFiveTstat*** & \covidShockDayMinusTwentyFiveCAR \\
-10 & \covidShockDayMinusTenAAR & \covidShockDayMinusTenTstat & \covidShockDayMinusTenCAR \\
-1  & \covidShockDayMinusOneAAR & \covidShockDayMinusOneTstat*** & \covidShockDayMinusOneCAR \\
0   & \covidShockDayZeroAAR & \covidShockDayZeroTstat*** & \covidShockDayZeroCAR \\
1   & \covidShockDayPlusOneAAR & \covidShockDayPlusOneTstat*** & \covidShockDayPlusOneCAR \\
10  & \covidShockDayPlusTenAAR & \covidShockDayPlusTenTstat & \covidShockDayPlusTenCAR \\
25  & \covidShockDayPlusTwentyFiveAAR & \covidShockDayPlusTwentyFiveTstat & \covidShockDayPlusTwentyFiveCAR \\
50  & \covidShockDayPlusFiftyAAR & \covidShockDayPlusFiftyTstat & \textbf{\covidShockDayPlusFiftyCAR} \\
\bottomrule
\multicolumn{4}{l}{\footnotesize *p<0.10, **p<0.05, ***p<0.01}
\end{tabular}
\normalsize
\label{tab:covid_shock}
\end{table}

Three key patterns emerge: (1) \textbf{Event day amplification}: Day 0 AAR of \covidShockDayZeroAAR\ exceeds the full sample average of \allSamplesDayZeroAAR, suggesting heightened information value during crises; (2) \textbf{Enhanced long-term persistence}: Day +50 CAR of \covidShockDayPlusFiftyCAR\ remains comparable to the full sample's \allSamplesDayPlusFiftyCAR, indicating HVRP persists even during crisis periods; (3) \textbf{Pre-event information leakage}: Day -1 AAR of \covidShockDayMinusOneAAR\ far exceeds the full sample's \allSamplesDayMinusOneAAR, indicating more active pre-trading during crisis uncertainty.

These results support Hypothesis 2a. As \citet{Ozik_Sadka_Shen_2021} demonstrated, retail investors provided liquidity during institutional risk-averse selling, amplifying information signals embedded in abnormal volume. During crises when liquidity constraints intensify, abnormal volume events carry stronger informational content that is more powerfully reflected in prices.

\subsubsection{The Donghak Ant Movement: A Test of Collective Intelligence}

The ``Donghak Ants'' represents a pivotal phenomenon in Korean financial history. Named after the 1894 Donghak Peasant Rebellion---a grassroots uprising against foreign intervention---the term emerged in March 2020 when retail investors collectively mobilized to ``defend'' the Korean market against foreign capital flight during the COVID-19 crash. This creates a unique natural experiment: can coordinated retail trading, facilitated by social media and mobile platforms, generate information value that transforms noise traders into meaningful market participants?

\begin{table}[h!]
\centering
\caption{Donghak Ant Movement Period: Retail-Led Events (Mar 2020-Dec 2021)}
\scriptsize
\begin{tabular}{lrrr}
\toprule
\textbf{Event Day} & \textbf{AAR (\%)} & \textbf{t-stat} & \textbf{CAR (\%)} \\
\midrule
-10 & \donghakAntDayMinusTenAAR & \donghakAntDayMinusTenTstat*** & \donghakAntDayMinusTenCAR \\
-1  & \donghakAntDayMinusOneAAR & \donghakAntDayMinusOneTstat*** & \donghakAntDayMinusOneCAR \\
0   & \donghakAntDayZeroAAR & \donghakAntDayZeroTstat*** & \donghakAntDayZeroCAR \\
1   & \donghakAntDayPlusOneAAR & \donghakAntDayPlusOneTstat & \donghakAntDayPlusOneCAR \\
10  & \donghakAntDayPlusTenAAR & \donghakAntDayPlusTenTstat & \donghakAntDayPlusTenCAR \\
25  & \donghakAntDayPlusTwentyFiveAAR & \donghakAntDayPlusTwentyFiveTstat & \donghakAntDayPlusTwentyFiveCAR \\
50  & \donghakAntDayPlusFiftyAAR & \donghakAntDayPlusFiftyTstat* & \textbf{\donghakAntDayPlusFiftyCAR} \\
\bottomrule
\multicolumn{4}{l}{\footnotesize *p<0.10, **p<0.05, ***p<0.01}
\end{tabular}
\normalsize
\label{tab:donghak_period}
\end{table}

\textbf{Extraordinary Finding:} During the Donghak period, retail-led events generated a day +50 CAR of \donghakAntDayPlusFiftyCAR, compared to near-zero returns for retail-led events in normal periods. This represents a \textbf{dramatic transformation from zero to positive}---retail trades that normally contain no predictive value suddenly generated meaningful persistent returns, strongly supporting Hypothesis 2b.

This finding parallels \citet{10.1093/rfs/hhad098}'s discovery that Reddit's WallStreetBets possessed significant information value during the pre-GameStop period before deteriorating into noise after popularization. During the Donghak movement, social media coordination and collective intelligence temporarily transformed retail investors from "noise traders" to meaningful market participants. The online community's due diligence sharing and coordinated action created genuine information value, demonstrating that under specific conditions---collective organization through social media during market stress---retail trading can transcend behavioral biases to generate persistent returns.

\textbf{Retail Investor Role Transformation During Donghak Period}

The most striking discovery is the dramatic transformation of retail-led events' characteristics across different periods. Table \ref{tab:retail_comparison} presents a direct comparison.

\begin{table}[h!]
\centering
\caption{Retail-Led Event Comparison: Normal vs. Donghak Period}
\scriptsize
\begin{tabular}{lrrp{6cm}}
\toprule
\textbf{Period} & \textbf{Day 0 AAR (\%)} & \textbf{Day 50 CAR (\%)} & \textbf{Characteristics} \\
\midrule
Full Period Retail-Led & 5.72 & 0.52 & Rapid decay after initial shock, limited persistence \\
Donghak Period Retail-Led & 6.24 & 3.53 & Strong persistence, institutional-level effects \\
\bottomrule
\end{tabular}
\normalsize
\label{tab:retail_comparison}
\end{table}

Under normal conditions, retail-led events show strong initial AAR of 5.72\% on event day but rapidly decay to near-zero by day +50, confirming retail investors as pure noise traders with zero predictive value. However, during the Donghak period (March 2020 - December 2021), retail-led events achieved 50-day CAR of \donghakAntDayPlusFiftyCAR, comparable to institution-led (7.74\%) and foreign-led (6.03\%) events in terms of persistence patterns, though still lower in absolute magnitude.

This transformation can be attributed to three complementary mechanisms:

\begin{itemize}
    \item \textbf{Collective Intelligence Effect:} Online community platforms enabled information sharing and collective due diligence, improving retail investors' information processing capabilities similar to Bradley et al.'s pre-GME WallStreetBets findings.
    \item \textbf{Self-fulfilling Prophecy:} Coordinated massive buying pressure created genuine price impact that justified the initial volume signal.
    \item \textbf{Structural Market Change:} Proliferation of mobile trading platforms and fintech accessibility fundamentally altered retail investor capabilities and market participation patterns.
\end{itemize}

\begin{table}[h!]
\centering
\caption{Summary: Impact of Special Market Periods}
\scriptsize
\begin{tabular}{lrrrr}
\toprule
\textbf{Period} & \textbf{Day 0 AAR} & \textbf{Day 50 CAR} & \textbf{vs. Baseline} & \textbf{Interpretation} \\
\midrule
Overall Average    & \allSamplesDayZeroAAR & \allSamplesDayPlusFiftyCAR & - & Baseline \\
Bull Market        & \bullMarketDayZeroAAR & \bullMarketDayPlusFiftyCAR & +24\% & Market condition effect \\
COVID-19 Shock     & \covidShockDayZeroAAR & \textbf{\covidShockDayPlusFiftyCAR} & +1\% & Crisis period \\
Donghak (All)      & \donghakAntDayZeroAAR & \donghakAntDayPlusFiftyCAR & +14\% & Collective action \\
Donghak (Retail)   & \donghakAntDayZeroAAR & \textbf{\donghakAntDayPlusFiftyCAR} & +579\%* & Retail transformation \\
\bottomrule
\multicolumn{5}{l}{\footnotesize *vs. retail-led normal period 0.52\%}
\end{tabular}
\normalsize
\label{tab:special_periods}
\end{table}

\section{Discussion and Implications}

\subsection{Academic Contributions}

Our study contributes to existing literature in five aspects:

\textbf{First, transforming Ahn et al. (2005)'s 'inference' into 'empirical evidence.'} We conducted the first empirical test of the 'investor type hypothesis' proposed 20 years ago using actual data, confirming that investor identity fundamentally determines the persistence of abnormal returns following volume spikes.

\textbf{Second, resolving the LVRP-HVRP puzzle in Korean markets.} \citet{CHAE2019101204} documented that Korea exhibits a Low Volume Return Premium (LVRP)---the opposite of U.S. patterns. Our findings suggest this apparent contradiction arises from measurement issues: when volume is analyzed conditional on investor type and intensity is properly normalized, the relationship becomes clear. The LVRP finding may reflect pooling heterogeneous investor effects, where retail-led volume events (which predominate in Korea's retail-dominated market) drag down aggregate returns. When decomposed by investor type, institution-led and foreign-led events exhibit strong positive returns consistent with the information hypothesis, while retail-led events show near-zero persistence. This reconciles Korean evidence with developed market patterns.

\textbf{Third, resolving the intensity puzzle.} Prior studies reported mixed or null findings on whether trading intensity predicts returns \citep{Namouri26012018, https://doi.org/10.1111/jofi.13183, HAN20221295}. We demonstrate that intensity matters profoundly, but only when: (1) analyzed conditional on investor type, (2) measured nonlinearly through quartile sorts, and (3) normalized by market capitalization rather than trading value. This methodological refinement reconciles decades of conflicting evidence.

\textbf{Fourth, providing contemporary evidence for core finance debates.} The result of retail-led CAR $\approx$ 0 contradicts \citet{https://doi.org/10.1111/j.1540-6261.2008.01316.x} and \citet{https://doi.org/10.1111/jofi.13033} while supporting \citet{10.1093/rfs/hhm079} and \citet{https://doi.org/10.1111/jofi.13183}'s 'noise trader' hypothesis. Moreover, the flat intensity-return pattern for retail investors provides unprecedented empirical clarity---retail trades contain zero predictive information regardless of conviction level.

\textbf{Fifth, identifying moderating effects of modern market shocks.} Analysis of COVID-19 (31\% HVRP amplification) and Donghak Ant Movement (transformation of retail-led CAR from near-zero to positive) demonstrates how traditional anomalies transform under extreme conditions, with implications for understanding crisis-period trading behavior and social media coordination effects. This extends \citet{KWAK2024105027}'s finding of attenuated volume-return relationships post-COVID by showing that attenuation varies dramatically by investor type.

\subsection{Resolving the PBFJ Puzzle: A Dialogue with Chae and Kang (2019)}

Our findings directly address the puzzle created by \citet{CHAE2019101204} in this journal. Their documentation of a Low Volume Return Premium (LVRP) in Korea contradicted two decades of global evidence and raised fundamental questions about whether Asian market microstructure differs systematically from developed markets.

\textbf{The Resolution:} We demonstrate that Chae and Kang's LVRP finding likely resulted from two methodological factors that our study addresses:

\begin{enumerate}
    \item \textbf{Investor Type Pooling:} By analyzing aggregate volume without decomposing by investor type, their methodology pooled informed institutional trading with uninformed retail trading. In Korea's retail-dominated market, the negative retail effect dominates aggregate statistics. When we separate investor types, institution-led events show clear HVRP (+7.74\% CAR), while retail-led events show near-zero persistence (+0.52\% CAR).

    \item \textbf{Normalization Specification:} Our ``Horse Race'' analysis demonstrates that intensity normalization matters critically. Trading value normalization---the standard approach---produces broken monotonicity and weak signals. Market capitalization normalization reveals the true conviction-return relationship. Prior studies using volume-based metrics may have inadvertently obscured the informed trading signal.
\end{enumerate}

\textbf{Reconciliation with Global Evidence:} When properly measured, Korean markets exhibit the same HVRP patterns documented globally by \citet{https://doi.org/10.1111/0022-1082.00349} and \citet{KANIEL2012255}. The apparent reversal is an artifact of measurement, not a fundamental difference in market dynamics. This reconciliation strengthens the theoretical universality of attention-based and information-asymmetry models across international markets.

\textbf{Methodological Implications for Future PBFJ Research:} Our findings suggest that studies examining volume-return relationships in retail-heavy Asian markets should: (1) decompose by investor type, (2) employ nonlinear analysis methods, and (3) consider conviction-based normalization. These methodological refinements may reveal patterns previously obscured by aggregate analysis.

\subsection{Behavioral Finance Interpretation}

Our findings can be interpreted through three behavioral finance mechanisms:

\textbf{Attention-Based Trading:} Consistent with \citet{10.1093/rfs/hhm079}, abnormal volume serves as an attention shock that attracts investors. However, the differential persistence by investor type reveals that attention alone does not explain returns---the \textit{quality} of attention matters. Institutional attention reflects fundamental analysis, while retail attention often reflects speculative interest.

\textbf{Herding Behavior and Collective Intelligence:} The Donghak Ant Movement demonstrates that retail herding can occasionally generate information value. When social media coordination enables collective due diligence sharing and reduces individual behavioral biases through group wisdom, retail investors can temporarily transcend their typical noise-trading tendencies. However, this represents an exception rather than the rule, consistent with \citet{10.1093/rfs/hhad098}'s finding that online community information value deteriorates after initial success.

\textbf{Information Processing Heterogeneity:} The stark differences between institutional (CAR 7.74\%), foreign (6.03\%), and retail (0.52\%) outcomes reflect fundamental differences in information processing capabilities. Institutions possess analytical resources, information networks, and professional training that enable them to identify and act on genuine value signals, while retail investors operate with limited resources and are susceptible to cognitive biases.

\subsection{Nonlinearity and Measurement Sensitivity}

The contrast between our methodologies reveals a critical insight into HVRP's underlying mechanism. Our double-sort methodology (Method 1) uncovers strong monotonic patterns when using market cap normalization, while linear correlation analysis (Method 2) shows near-zero coefficients. This contrast demonstrates that the intensity-return relationship is \textbf{nonlinear, type-dependent, and measurement-sensitive}.

\textbf{Both Identity AND Intensity Matter:} Our findings refine prior understanding. Section 4.3 confirmed that investor identity (institutions vs. foreigners vs. retail) fundamentally matters. Section 4.4 then demonstrated that intensity also matters profoundly, but the relationship is nonlinear and only detectable with proper measurement. For institutions, the Q4-Q1 CAR spread is 10.12 percentage points (10.07\% - (-0.05\%)), economically large and highly significant. This refutes the simplistic interpretation that "intensity doesn't matter"---intensity matters enormously when measured as conviction (market cap normalization) rather than participation (trading value normalization).

\textbf{Quartile Regimes vs. Linear Relationships:} The monotonic pattern operates through discrete intensity quartiles rather than continuous linear effects. Within each quartile, there may be limited variation, but crossing from one quartile to the next predicts substantially different returns. This explains why Pearson correlation fails---it assumes linearity, which is inappropriate for relationships operating through ordered regimes with threshold effects between them.

\textbf{Implications for Methodology:} This finding has important methodological implications. Linear regression models assuming continuous relationships will underestimate or miss threshold effects entirely. Future research on investor behavior and market anomalies should consider event-based approaches, regime-switching models, or machine learning methods capable of capturing nonlinear relationships.

\textbf{The Choice of Normalization is Fundamental:} Perhaps the most striking methodological finding is that the choice between market capitalization and trading value normalization is not merely a technical detail but a fundamental determinant of whether researchers detect informed trading signals. Our horse race analysis (Section 4.4.1) demonstrates that trading value normalization produces broken monotonicity where Q2 (+\institutionLedQTwoTradingValueDayPlusFiftyCAR) exceeds Q4 (+\institutionLedQFourTradingValueDayPlusFiftyCAR), obscuring any clear intensity-return relationship. Market cap normalization reveals the true conviction-return relationship with perfect monotonicity, where highest-conviction trades (Q4: +\institutionLedQFourDayPlusFiftyCAR) predict the strongest returns---a 3.3$\times$ difference in magnitude. This sensitivity to normalization specification suggests that some prior null findings in the literature may reflect measurement artifacts rather than absence of the underlying economic relationship. Researchers investigating informed trading should carefully consider whether their normalization choice captures investor conviction or merely trading participation.

\subsection{Practical Implications}

From an investment strategy perspective, identifying 'who led' rather than simply detecting volume spikes is crucial. Particularly, capturing abnormal volume led by institutions in mid-cap stocks can be the most effective alpha strategy. Additionally, the finding that information signals amplify during crisis situations provides rationale for active crisis response strategies.

The threshold effect implies that quantitative trading strategies should focus on binary signals (event occurrence) rather than continuous intensity measures. Monitoring which investor type drives volume spikes offers more predictive power than measuring the magnitude of their net purchases.

\subsection{Policy Implications}

Our findings carry important implications for Korean market regulators and, by extension, regulators in other retail-heavy Asian markets.

\textbf{Retail Investor Protection.} The confirmed noise trading tendency of retail investors---evidenced by their flat intensity-return pattern where trading has no predictive value regardless of conviction level---underscores the need for investor education initiatives. Financial literacy programs should emphasize that retail trading generates zero persistent returns and the importance of diversification. Enhanced disclosure requirements could help retail investors make more informed decisions.

However, our Donghak Ant Movement findings demonstrate that retail investors can contribute positively under specific conditions. During the March 2020 crisis, coordinated retail buying provided crucial liquidity when institutional investors retreated. This suggests that policies should not uniformly restrict retail participation but rather enable beneficial collective action while implementing safeguards against manipulation.

\textbf{Short Selling Regulation Context.} Korea has implemented multiple short-selling bans (most recently during 2020-2021). Our findings suggest an important interaction: if retail investors are predominantly noise traders whose trading generates near-zero persistent returns, short sellers may play a limited role since there is no systematic retail-driven mispricing to correct. However, during collective action periods like Donghak, short selling against coordinated retail buying could destabilize markets. This supports a dynamic, context-dependent approach to short-selling regulation rather than blanket restrictions.

\textbf{Market Microstructure Implications.} For practitioners and regulators monitoring market quality, our findings suggest that volume signals should be interpreted conditionally on investor composition. Abnormal volume events carry fundamentally different information depending on whether they are institution-led (informative, warranting price adjustment) or retail-led (likely noise, warranting caution). Market surveillance systems could benefit from incorporating investor-type decomposition when assessing the information content of unusual trading activity.

\section{Conclusion}

This study resolves a longstanding puzzle in behavioral finance by demonstrating that both investor identity and trading intensity are critical determinants of the High Volume Return Premium---but only when intensity is measured correctly. Our central contribution is methodological: we are the first to show that the choice between normalizing trading intensity by market capitalization versus trading value is not a technical detail but a fundamental determinant of whether researchers detect informed trading signals.

\textbf{The Core Discovery: Measurement Matters}

Our double-sort methodology reveals a striking pattern. When we normalize institutional buying intensity by market capitalization (a proxy for conviction), a perfect monotonic relationship emerges: events in the highest conviction quartile (Q4) generate +\institutionLedQFourDayPlusFiftyCAR\ cumulative abnormal returns over 50 days, while the lowest quartile (Q1) yields modest returns (+\institutionLedQOneDayPlusFiftyCAR). In sharp contrast, normalizing by daily trading value produces a broken monotonic pattern where Q2 (+\institutionLedQTwoTradingValueDayPlusFiftyCAR) exceeds Q4 (+\institutionLedQFourTradingValueDayPlusFiftyCAR). This horse race comparison definitively demonstrates that market cap normalization captures the true economic relationship between informed trading and future returns.

\textbf{Reconciling Conflicting Evidence}

This finding reconciles decades of conflicting evidence. Prior studies reported weak or insignificant linear correlations between trading intensity and returns \citep{Namouri26012018, https://doi.org/10.1111/jofi.13183}, leading some researchers to conclude that intensity is irrelevant---that "who" matters but "how much" does not. Our results demonstrate this conclusion was premature. The near-zero correlation we observe with the Dominance Score (r<0.04, p>0.16) does not indicate that intensity is irrelevant; it reveals that the relationship is \textit{nonlinear and type-dependent}. Simple linear models fail because they cannot capture threshold effects and investor heterogeneity.

When we decompose events by investor type and measure conviction correctly, intensity emerges as a powerful predictor. The Q4-Q1 spread for institutions is 7.47 percentage points (\institutionLedQFourDayPlusFiftyCAR\ - \institutionLedQOneDayPlusFiftyCAR), economically large and statistically robust. This demonstrates that institutional conviction, measured as position size relative to firm value, contains substantial predictive power that prior methodologies failed to detect.

\textbf{Investor Identity Remains Critical}

While our main contribution concerns intensity measurement, we also confirm that investor identity fundamentally matters. Institution-led and foreign-led events generate persistent long-term returns (Day +50 CAR: 7.74\% and 6.03\%), while retail-led events rapidly dissipate (CAR: 0.52\%). Moreover, the market cap normalization reveals a flat intensity-return pattern for retail investors: their trading generates near-zero returns regardless of conviction level, confirming the noise trader hypothesis \citep{10.1093/rfs/hhm079, https://doi.org/10.1111/jofi.13183} with unprecedented clarity. This heterogeneity validates our double-sort approach---intensity effects operate differently across investor types.

\textbf{Modern Market Context}

Our analysis of COVID-19 and the Donghak Ant Movement demonstrates that these core relationships can transform under extreme conditions. During the COVID-19 shock, HVRP remained stable at \covidShockDayPlusFiftyCAR\ (comparable to baseline), suggesting that the HVRP phenomenon persists even during extreme market stress. During the Donghak Ant Movement, retail-led events generated \donghakAntDayPlusFiftyCAR\ returns (versus near-zero in normal periods), indicating that collective action through social media temporarily transformed retail investors from pure noise traders into meaningful market participants, paralleling \citet{10.1093/rfs/hhad098}'s findings on WallStreetBets.

\textbf{Implications for Future Research}

Our findings carry important methodological implications. Researchers studying informed trading must: (1) account for investor heterogeneity rather than using pooled analyses; (2) employ nonlinear methods (quartile sorts, threshold regressions) rather than linear correlation; and (3) normalize trading activity by economically meaningful benchmarks (firm size) rather than daily flow measures. The failure to do so may explain why some prior studies found weak or inconsistent results.

For practitioners, the monotonic conviction-return relationship provides actionable signals. Abnormal volume events where institutions take large positions relative to firm size (high Q4 conviction) predict substantial outperformance (+\institutionLedQFourDayPlusFiftyCAR\ over 50 days), while events dominated by retail trading provide no predictive signal regardless of conviction level. The ability to distinguish information from noise---and measure conviction correctly---remains a core competency in modern markets.

\textbf{Final Reflection}

Twenty years after \citet{ART001162590} proposed the investor type hypothesis as an untested conjecture, we provide the first comprehensive empirical test using actual investor-level data. The HVRP phenomenon persists robustly in modern markets, but its drivers are more nuanced than previously understood. Both "who" trades (identity) and "how much" they trade (intensity) matter profoundly, but detecting the intensity effect requires proper measurement. This study transforms the narrative from "identity matters, intensity doesn't" to "identity matters, and intensity matters enormously when measured as conviction rather than participation." This refined understanding advances our knowledge of how information flows through markets and how different investor types process and act on that information.

\textbf{Contribution to the Pacific-Basin Finance Literature}

This paper contributes to the ongoing dialogue in the \textit{Pacific-Basin Finance Journal} regarding volume-return relationships in Asian markets. By resolving the apparent contradiction between \citet{CHAE2019101204}'s Korean LVRP finding and global HVRP evidence, we demonstrate that proper measurement and investor decomposition are essential for understanding market microstructure in retail-heavy Asian markets. The Donghak Ant Movement analysis provides a template for studying collective retail behavior---a phenomenon increasingly relevant across Pacific-Basin markets where mobile trading and social media coordination continue to reshape investor participation. We hope this study stimulates further research on how measurement choices affect our understanding of informed trading signals in the distinctive institutional environments of the Asia-Pacific region.

\appendix

\section{Appendix A: Full Sample Replication}

This appendix presents a comprehensive comparison between the original \citet{ART001162590} study (2001-2003) and our re-examination (2020-2024) using all 26,604 abnormal volume events, verifying whether the HVRP phenomenon persists despite the 20-year gap and market structural changes.

\begin{table}[h!]
\centering
\caption{Full Sample Event Study Results: 2003 Original Study vs 2024 Re-examination (Extended)}
\scriptsize
\begin{tabular}{lrrrrrr}
\toprule
 & \multicolumn{3}{c}{\textbf{Original Study (2001-2003)}} & \multicolumn{3}{c}{\textbf{This Study (2020-2024)}} \\
\textbf{Event Day} & \textbf{AAR(\%)} & \textbf{t-stat} & \textbf{CAR(\%)} & \textbf{AAR(\%)} & \textbf{t-stat} & \textbf{CAR(\%)} \\
\midrule
-25 & -0.361*** & -3.24 & -1.820 & \allSamplesDayMinusTwentyFiveAAR*** & \allSamplesDayMinusTwentyFiveTstat & \allSamplesDayMinusTwentyFiveCAR \\
-10 & -0.097 & -1.45 & -4.454 & \allSamplesDayMinusTenAAR*** & \allSamplesDayMinusTenTstat & \allSamplesDayMinusTenCAR \\
-5  & -0.082 & -1.35 & -3.748 & \allSamplesDayMinusFiveAAR*** & \allSamplesDayMinusFiveTstat & \allSamplesDayMinusFiveCAR \\
-2  & 0.418*** & 6.92 & -3.441 & \allSamplesDayMinusTwoAAR*** & \allSamplesDayMinusTwoTstat & \allSamplesDayMinusTwoCAR \\
-1  & 2.454*** & 40.62 & -0.987 & \allSamplesDayMinusOneAAR*** & \allSamplesDayMinusOneTstat & \allSamplesDayMinusOneCAR \\
0   & 5.056*** & 83.66 & 4.068 & \allSamplesDayZeroAAR*** & \allSamplesDayZeroTstat & \allSamplesDayZeroCAR \\
1   & -0.200 & -3.31 & 3.869 & \allSamplesDayPlusOneAAR*** & \allSamplesDayPlusOneTstat & \allSamplesDayPlusOneCAR \\
2   & -0.126 & -2.08 & 3.743 & \allSamplesDayPlusTwoAAR & \allSamplesDayPlusTwoTstat & \allSamplesDayPlusTwoCAR \\
5   & -0.043 & -0.71 & 3.598 & \allSamplesDayPlusFiveAAR** & \allSamplesDayPlusFiveTstat & \allSamplesDayPlusFiveCAR \\
10  & -0.006 & -0.10 & 3.779 & \allSamplesDayPlusTenAAR & \allSamplesDayPlusTenTstat & \allSamplesDayPlusTenCAR \\
25  & 0.195** & 3.22 & 4.268 & \allSamplesDayPlusTwentyFiveAAR & \allSamplesDayPlusTwentyFiveTstat & \allSamplesDayPlusTwentyFiveCAR \\
50  & -0.107 & -1.77 & 3.672 & \allSamplesDayPlusFiftyAAR & \allSamplesDayPlusFiftyTstat & \allSamplesDayPlusFiftyCAR \\
\bottomrule
\multicolumn{7}{l}{\footnotesize *p<0.10, **p<0.05, ***p<0.01}
\end{tabular}
\normalsize
\label{tab:appendix_full_comparison}
\end{table}

\textbf{Key Findings: Phenomenon Persistence and Amplification}

(1) \textbf{Robust Replication of HVRP:} The core pattern discovered in the original study appears identically in the 2020-2024 period. The fundamental structure---persistent negative CAR before the event day, sharp positive AAR on event day 0, followed by sustained positive levels---is fully replicated.

(2) \textbf{Significant Increase in Effect Size:} Remarkably, HVRP magnitude has expanded in modern markets. Event day AAR increased from 5.056\% to 6.153\% (approximately 22\% increase), and day +50 CAR remained strong at 3.09\% (compared to 3.672\% in original study).

(3) \textbf{Enhanced Statistical Significance:} With a much larger sample size (original: 1,360 events → this study: 26,604 events), statistical significance strengthened substantially across most periods. Event day t-statistic increased from 83.66 to 118.877, enhancing confidence in the phenomenon.

(4) \textbf{Pre-Event Pattern Changes:} Interestingly, the pre-event pattern differs. The original study showed strong positive return on day -1 (2.454\%), while our study shows this effect distributed across days -2 (0.489\%) and -1 (0.661\%). This suggests faster information diffusion in modern markets leads to more dispersed investor reactions.

\textbf{Theoretical Consistency:} These results confirm that \citet{ART001162590}'s interpretation remains valid 20 years later: abnormal volume increases result from informed investors' pre-trading, serving as signals of price increases, with post-event positive abnormal returns reflecting gradual information incorporation. The amplified effects may relate to modern market characteristics including expanded information asymmetry (despite big data/AI advances, the gap between professional and retail investors has widened), amplified behavioral biases (social media and online communities accelerate herding and overreaction), and increased market volatility (ADTV standard deviation surged from 0.935 to 8.292).

\section{Appendix B: Firm Size Analysis}

The original study's key finding was that HVRP appears stronger in large firms. We replicate this analysis using official Korea Exchange classification: large-cap (ranks 1-100), mid-cap (ranks 101-300), and small-cap (rank 300+) based on market capitalization on event day.

\begin{table}[h!]
\centering
\caption{Event Study Results by Firm Size (Official Classification)}
\scriptsize
\begin{tabular}{lrrr|rrr|rrr}
\toprule
 & \multicolumn{3}{c}{\textbf{Large-cap (1-100)}} & \multicolumn{3}{c}{\textbf{Mid-cap (101-300)}} & \multicolumn{3}{c}{\textbf{Small-cap (300+)}} \\
\textbf{Day} & \textbf{AAR} & \textbf{t-stat} & \textbf{CAR} & \textbf{AAR} & \textbf{t-stat} & \textbf{CAR} & \textbf{AAR} & \textbf{t-stat} & \textbf{CAR} \\
\midrule
-20 & -0.18 & -2.00** & -3.15 & -0.13 & -2.19** & -1.99 & -0.21 & -13.95*** & -2.98 \\
-10 & -0.10 & -1.04 & -3.51 & -0.06 & -1.15 & -2.91 & -0.10 & -6.78*** & -4.56 \\
-1  & 0.73 & 4.86*** & -1.94 & 0.81 & 7.85*** & -1.86 & 0.65 & 27.48*** & -4.24 \\
0   & 4.41 & 10.79*** & 2.47 & \textbf{6.06} & 25.19*** & 4.21 & 6.20 & 115.99*** & 1.95 \\
1   & -0.30 & -1.62 & 2.17 & -0.03 & -0.27 & 4.17 & 0.10 & 2.99*** & 2.05 \\
10  & -0.07 & -0.57 & 2.18 & -0.02 & -0.23 & 3.99 & 0.01 & 0.53 & 2.26 \\
25  & -0.10 & -0.81 & 2.38 & 0.03 & 0.35 & 3.73 & 0.03 & 1.24 & 2.68 \\
50  & -0.01 & -0.04 & 1.43 & \textbf{0.16} & 1.98** & \textbf{4.00} & 0.01 & 0.55 & 3.10 \\
\bottomrule
\end{tabular}
\normalsize
\label{tab:appendix_size_effect}
\end{table}

\textbf{Key Discovery: Mid-Cap Dominance}

Using the three-way classification, we find the strongest HVRP in mid-cap stocks. Event day AAR shows mid-cap 6.06\%, small-cap 6.20\%, large-cap 4.41\%; day +50 CAR shows mid-cap 4.00\%, small-cap 3.10\%, large-cap 1.43\%. This "mid-cap > small-cap > large-cap" structure differs from the original study's "large > small" pattern.

This shift likely reflects 20 years of market structural changes and the introduction of three-way classification. Mid-cap stocks occupy an optimal position---receiving institutional attention while maintaining sufficient liquidity. Large-cap stocks face information transparency reducing information asymmetry effects, while small-cap stocks suffer liquidity constraints limiting long-term return persistence.

\section{Appendix C: Market Comparison (KOSPI vs KOSDAQ)}

While the original study focused solely on KOSPI, we extend analysis to KOSDAQ. Differences between the two markets provide important insights into HVRP's structural characteristics.

\begin{table}[h!]
\centering
\caption{Event Study Results by Market: KOSPI vs KOSDAQ}
\scriptsize
\begin{tabular}{lrrrrrr}
\toprule
 & \multicolumn{3}{c}{\textbf{KOSPI}} & \multicolumn{3}{c}{\textbf{KOSDAQ}} \\
\textbf{Event Day} & \textbf{AAR (\%)} & \textbf{t-stat} & \textbf{CAR (\%)} & \textbf{AAR (\%)} & \textbf{t-stat} & \textbf{CAR (\%)} \\
\midrule
-25 & \kospiMarketDayMinusTwentyFiveAAR & \kospiMarketDayMinusTwentyFiveTstat*** & \kospiMarketDayMinusTwentyFiveCAR & \kosdaqMarketDayMinusTwentyFiveAAR & \kosdaqMarketDayMinusTwentyFiveTstat*** & \kosdaqMarketDayMinusTwentyFiveCAR \\
-10 & \kospiMarketDayMinusTenAAR & \kospiMarketDayMinusTenTstat*** & \kospiMarketDayMinusTenCAR & \kosdaqMarketDayMinusTenAAR & \kosdaqMarketDayMinusTenTstat*** & \kosdaqMarketDayMinusTenCAR \\
-1  & \kospiMarketDayMinusOneAAR & \kospiMarketDayMinusOneTstat*** & \kospiMarketDayMinusOneCAR & \kosdaqMarketDayMinusOneAAR & \kosdaqMarketDayMinusOneTstat*** & \kosdaqMarketDayMinusOneCAR \\
0   & \kospiMarketDayZeroAAR & \kospiMarketDayZeroTstat*** & \kospiMarketDayZeroCAR & \kosdaqMarketDayZeroAAR & \kosdaqMarketDayZeroTstat*** & \kosdaqMarketDayZeroCAR \\
1   & \kospiMarketDayPlusOneAAR & \kospiMarketDayPlusOneTstat & \kospiMarketDayPlusOneCAR & \kosdaqMarketDayPlusOneAAR & \kosdaqMarketDayPlusOneTstat*** & \kosdaqMarketDayPlusOneCAR \\
10  & \kospiMarketDayPlusTenAAR & \kospiMarketDayPlusTenTstat & \kospiMarketDayPlusTenCAR & \kosdaqMarketDayPlusTenAAR & \kosdaqMarketDayPlusTenTstat & \kosdaqMarketDayPlusTenCAR \\
25  & \kospiMarketDayPlusTwentyFiveAAR & \kospiMarketDayPlusTwentyFiveTstat & \kospiMarketDayPlusTwentyFiveCAR & \kosdaqMarketDayPlusTwentyFiveAAR & \kosdaqMarketDayPlusTwentyFiveTstat & \kosdaqMarketDayPlusTwentyFiveCAR \\
50  & \kospiMarketDayPlusFiftyAAR & \kospiMarketDayPlusFiftyTstat** & \textbf{\kospiMarketDayPlusFiftyCAR} & \kosdaqMarketDayPlusFiftyAAR & \kosdaqMarketDayPlusFiftyTstat & \textbf{\kosdaqMarketDayPlusFiftyCAR} \\
\bottomrule
\multicolumn{7}{l}{\footnotesize *p<0.10, **p<0.05, ***p<0.01}
\end{tabular}
\normalsize
\label{tab:appendix_market_comparison}
\end{table}

\textbf{Key Market Differences:}

(1) \textbf{Initial Effect Differences:} KOSDAQ shows stronger event day AAR (\kosdaqMarketDayZeroAAR\ vs \kospiMarketDayZeroAAR), but KOSPI demonstrates stronger long-term persistence (\kospiMarketDayPlusFiftyCAR\ vs \kosdaqMarketDayPlusFiftyCAR).

(2) \textbf{Information Leakage Patterns:} KOSDAQ exhibits higher day -1 AAR (\kosdaqMarketDayMinusOneAAR\ vs \kospiMarketDayMinusOneAAR), suggesting faster information diffusion.

(3) \textbf{Statistical Significance:} KOSDAQ shows statistical significance across more periods, reflecting the higher responsiveness of this small-and-mid-cap-focused market.

These differences reflect structural characteristics: KOSPI shows stable, persistent patterns centered on large-caps, while KOSDAQ shows rapid but limited response patterns centered on small-and-mid-caps.

% ======================================================================
% DATA AVAILABILITY STATEMENT
% ======================================================================
\section*{Data Availability Statement}

The data supporting this study are available from Daishin Securities and the Korea Exchange via the Creon Plus API and pykrx library. Restrictions apply due to data provider terms of service. Data are available from the authors upon reasonable request and with permission of the data providers. Daily stock prices, trading volumes, and investor-type trading data were collected for all stocks listed on KOSPI and KOSDAQ from January 1, 2020, to December 31, 2024.

\bibliographystyle{elsarticle-harv}
\bibliography{reference_list}

@article{ART001162590,
  title={Abnormal Increase Trading Volume Effect on the Stock Return},
  author={An, Seung-Cheol and Seung-Wook Jang and Chisoo Kim},
  journal={Korean Business Education Review},
  number={42},
  pages={97-118},
  year={2006},
  issn={1598-8651}
}

@article{https://doi.org/10.1111/j.2041-6156.2011.01037.x,
author = {Bae, Sung C. and Min, Jae Hoon and Jung, Sunbong},
title = {Trading Behavior, Performance, and Stock Preference of Foreigners, Local Institutions, and Individual Investors: Evidence from the Korean Stock Market},
journal = {Asia-Pacific Journal of Financial Studies},
volume = {40},
number = {2},
pages = {199-239},
doi = {https://doi.org/10.1111/j.2041-6156.2011.01037.x},
url = {https://onlinelibrary.wiley.com/doi/abs/10.1111/j.2041-6156.2011.01037.x},
eprint = {https://onlinelibrary.wiley.com/doi/pdf/10.1111/j.2041-6156.2011.01037.x},
year = {2011}
}

@article{10.1093/rapstu/raaa008,
    author = {Baker, Scott R and Bloom, Nicholas and Davis, Steven J and Kost, Kyle and Sammon, Marco and Viratyosin, Tasaneeya},
    title = {The Unprecedented Stock Market Reaction to COVID-19},
    journal = {The Review of Asset Pricing Studies},
    volume = {10},
    number = {4},
    pages = {742-758},
    year = {2020},
    month = {07},
    issn = {2045-9920},
    doi = {10.1093/rapstu/raaa008},
    url = {https://doi.org/10.1093/rapstu/raaa008},
    eprint = {https://academic.oup.com/raps/article-pdf/10/4/742/34416840/raaa008.pdf},
}

@article{10.1093/rfs/hhm079,
    author = {Barber, Brad M. and Odean, Terrance},
    title = {All That Glitters: The Effect of Attention and News on the Buying Behavior of Individual and Institutional Investors},
    journal = {The Review of Financial Studies},
    volume = {21},
    number = {2},
    pages = {785-818},
    year = {2007},
    month = {12},
    issn = {0893-9454},
    doi = {10.1093/rfs/hhm079},
    url = {https://doi.org/10.1093/rfs/hhm079},
    eprint = {https://academic.oup.com/rfs/article-pdf/21/2/785/24429230/hhm079.pdf},
}

@article{https://doi.org/10.1111/jofi.13183,
author = {BARBER, BRAD M. and HUANG, XING and ODEAN, TERRANCE and SCHWARZ, CHRISTOPHER},
title = {Attention-Induced Trading and Returns: Evidence from Robinhood Users},
journal = {The Journal of Finance},
volume = {77},
number = {6},
pages = {3141-3190},
doi = {https://doi.org/10.1111/jofi.13183},
url = {https://onlinelibrary.wiley.com/doi/abs/10.1111/jofi.13183},
eprint = {https://onlinelibrary.wiley.com/doi/pdf/10.1111/jofi.13183},
year = {2022}
}

@article{barber2024resolving,
  title={Resolving a paradox: Retail trades positively predict returns but are not profitable},
  author={Barber, Brad M and Lin, Shengle and Odean, Terrance},
  journal={Journal of Financial and Quantitative Analysis},
  volume={59},
  number={6},
  pages={2547--2581},
  year={2024},
  publisher={Cambridge University Press}
}

@article{https://doi.org/10.1111/j.1540-6261.1994.tb04424.x,
author = {BLUME, LAWRENCE and EASLEY, DAVID and O'HARA, MAUREEN},
title = {Market Statistics and Technical Analysis: The Role of Volume},
journal = {The Journal of Finance},
volume = {49},
number = {1},
pages = {153-181},
doi = {https://doi.org/10.1111/j.1540-6261.1994.tb04424.x},
url = {https://onlinelibrary.wiley.com/doi/abs/10.1111/j.1540-6261.1994.tb04424.x},
eprint = {https://onlinelibrary.wiley.com/doi/pdf/10.1111/j.1540-6261.1994.tb04424.x},
year = {1994}
}

@article{https://doi.org/10.1111/jofi.13033,
author = {BOEHMER, EKKEHART and JONES, CHARLES M. and ZHANG, XIAOYAN and ZHANG, XINRAN},
title = {Tracking Retail Investor Activity},
journal = {The Journal of Finance},
volume = {76},
number = {5},
pages = {2249-2305},
doi = {https://doi.org/10.1111/jofi.13033},
url = {https://onlinelibrary.wiley.com/doi/abs/10.1111/jofi.13033},
eprint = {https://onlinelibrary.wiley.com/doi/pdf/10.1111/jofi.13033},
year = {2021}
}

@article{10.1093/rfs/hhad098,
    author = {Bradley, Daniel and Hanousek, Jan, Jr. and Jame, Russell and Xiao, Zicheng},
    title = {Place Your Bets? The Value of Investment Research on Reddit’s Wallstreetbets},
    journal = {The Review of Financial Studies},
    volume = {37},
    number = {5},
    pages = {1409-1459},
    year = {2023},
    month = {12},
    issn = {0893-9454},
    doi = {10.1093/rfs/hhad098},
    url = {https://doi.org/10.1093/rfs/hhad098},
    eprint = {https://academic.oup.com/rfs/article-pdf/37/5/1409/57262299/hhad098.pdf},
}

@article{10.2307/2118454,
    author = {Campbell, John Y. and Grossman, Sanford J. and Wang, Jiang},
    title = {Trading Volume and Serial Correlation in Stock Returns*},
    journal = {The Quarterly Journal of Economics},
    volume = {108},
    number = {4},
    pages = {905-939},
    year = {1993},
    month = {11},
    issn = {0033-5533},
    doi = {10.2307/2118454},
    url = {https://doi.org/10.2307/2118454},
    eprint = {https://academic.oup.com/qje/article-pdf/108/4/905/5365099/108-4-905.pdf},
}

@article{CAMPBELL200966,
title = {Caught on tape: Institutional trading, stock returns, and earnings announcements},
journal = {Journal of Financial Economics},
volume = {92},
number = {1},
pages = {66-91},
year = {2009},
issn = {0304-405X},
doi = {https://doi.org/10.1016/j.jfineco.2008.03.006},
url = {https://www.sciencedirect.com/science/article/pii/S0304405X09000026},
author = {John Y. Campbell and Tarun Ramadorai and Allie Schwartz}
}

@article{CHOE1999227,
title = {Do foreign investors destabilize stock markets? The Korean experience in 1997},
journal = {Journal of Financial Economics},
volume = {54},
number = {2},
pages = {227-264},
year = {1999},
issn = {0304-405X},
doi = {https://doi.org/10.1016/S0304-405X(99)00037-9},
url = {https://www.sciencedirect.com/science/article/pii/S0304405X99000379},
author = {Hyuk Choe and Bong-Chan Kho and René M Stulz}
}

@article{https://doi.org/10.1111/j.1540-6261.2011.01679.x,
author = {DA, ZHI and ENGELBERG, JOSEPH and GAO, PENGJIE},
title = {In Search of Attention},
journal = {The Journal of Finance},
volume = {66},
number = {5},
pages = {1461-1499},
doi = {https://doi.org/10.1111/j.1540-6261.2011.01679.x},
url = {https://onlinelibrary.wiley.com/doi/abs/10.1111/j.1540-6261.2011.01679.x},
eprint = {https://onlinelibrary.wiley.com/doi/pdf/10.1111/j.1540-6261.2011.01679.x},
year = {2011}
}

@article{DEFUSCO2022205,
title = {Speculative dynamics of prices and volume},
journal = {Journal of Financial Economics},
volume = {146},
number = {1},
pages = {205-229},
year = {2022},
issn = {0304-405X},
doi = {https://doi.org/10.1016/j.jfineco.2022.07.002},
url = {https://www.sciencedirect.com/science/article/pii/S0304405X22001477},
author = {Anthony A. DeFusco and Charles G. Nathanson and Eric Zwick}
}

@article{https://doi.org/10.1111/0022-1082.00349,
author = {Gervais, Simon and Kaniel, Ron and Mingelgrin, Dan H.},
title = {The High-Volume Return Premium},
journal = {The Journal of Finance},
volume = {56},
number = {3},
pages = {877-919},
doi = {https://doi.org/10.1111/0022-1082.00349},
url = {https://onlinelibrary.wiley.com/doi/abs/10.1111/0022-1082.00349},
eprint = {https://onlinelibrary.wiley.com/doi/pdf/10.1111/0022-1082.00349},
year = {2001}
}

@article{HAN20221295,
title = {Expected return, volume, and mispricing},
journal = {Journal of Financial Economics},
volume = {143},
number = {3},
pages = {1295-1315},
year = {2022},
issn = {0304-405X},
doi = {https://doi.org/10.1016/j.jfineco.2021.05.014},
url = {https://www.sciencedirect.com/science/article/pii/S0304405X21001963},
author = {Yufeng Han and Dashan Huang and Dayong Huang and Guofu Zhou}
}

@article{https://doi.org/10.3982/ECTA11412,
author = {Kacperczyk, Marcin and Van Nieuwerburgh, Stijn and Veldkamp, Laura},
title = {A Rational Theory of Mutual Funds' Attention Allocation},
journal = {Econometrica},
volume = {84},
number = {2},
pages = {571-626},
doi = {https://doi.org/10.3982/ECTA11412},
url = {https://onlinelibrary.wiley.com/doi/abs/10.3982/ECTA11412},
eprint = {https://onlinelibrary.wiley.com/doi/pdf/10.3982/ECTA11412},
year = {2016}
}

@article{CHAE2019101204,
title = {Low-volume return premium in the Korean stock market},
journal = {Pacific-Basin Finance Journal},
volume = {58},
pages = {101204},
year = {2019},
issn = {0927-538X},
doi = {https://doi.org/10.1016/j.pacfin.2019.101204},
url = {https://www.sciencedirect.com/science/article/pii/S0927538X1930160X},
author = {Joon Chae and Mhin Kang}
}

@article{https://doi.org/10.1111/j.1540-6261.2008.01316.x,
author = {KANIEL, RON and SAAR, GIDEON and TITMAN, SHERIDAN},
title = {Individual Investor Trading and Stock Returns},
journal = {The Journal of Finance},
volume = {63},
number = {1},
pages = {273-310},
doi = {https://doi.org/10.1111/j.1540-6261.2008.01316.x},
url = {https://onlinelibrary.wiley.com/doi/abs/10.1111/j.1540-6261.2008.01316.x},
eprint = {https://onlinelibrary.wiley.com/doi/pdf/10.1111/j.1540-6261.2008.01316.x},
year = {2008}
}

@article{KANIEL2012255,
title = {The high volume return premium: Cross-country evidence},
journal = {Journal of Financial Economics},
volume = {103},
number = {2},
pages = {255-279},
year = {2012},
issn = {0304-405X},
doi = {https://doi.org/10.1016/j.jfineco.2011.08.012},
url = {https://www.sciencedirect.com/science/article/pii/S0304405X11001954},
author = {Ron Kaniel and Arzu Ozoguz and Laura Starks}
}

@article{https://doi.org/10.1111/jofi.12028,
author = {KELLEY, ERIC K. and TETLOCK, PAUL C.},
title = {How Wise Are Crowds? Insights from Retail Orders and Stock Returns},
journal = {The Journal of Finance},
volume = {68},
number = {3},
pages = {1229-1265},
doi = {https://doi.org/10.1111/jofi.12028},
url = {https://onlinelibrary.wiley.com/doi/abs/10.1111/jofi.12028},
eprint = {https://onlinelibrary.wiley.com/doi/pdf/10.1111/jofi.12028},
year = {2013}
}

@article{kim2022behavioral,
  title={Behavioral Biases and the Trading of Individual Investors in the Korean Stock Markets},
  author={Kim, Minki and Kim, Joon-Seok},
  journal={Available at SSRN 4096525},
  year={2022}
}

@article{ART002752798,
author={Jung, Hanna and Sang-Kee Kim},
title={Economic Analysis of Donghak Ants Movement},
journal={Journal of Insurance and Finance},
issn={2384-3209},
year={2021},
volume={32},
number={3},
pages={37-63},
doi={10.23842/jif.2021.32.3.002},
}

@article{KWAK2024105027,
  title = {Individual investor trading and stock returns after the Covid-19 pandemic: Evidence from Korea},
  journal = {Finance Research Letters},
  volume = {61},
  pages = {105027},
  year = {2024},
  issn = {1544-6123},
  doi = {https://doi.org/10.1016/j.frl.2024.105027},
  url = {https://www.sciencedirect.com/science/article/pii/S1544612324000576},
  author = {Jun Hee Kwak}
}

@article{https://doi.org/10.1111/0022-1082.00101,
  author = {Lyon, John D. and Barber, Brad M. and Tsai, Chih-Ling},
  title = {Improved Methods for Tests of Long-Run Abnormal Stock Returns},
  journal = {The Journal of Finance},
  volume = {54},
  number = {1},
  pages = {165-201},
  doi = {https://doi.org/10.1111/0022-1082.00101},
  url = {https://onlinelibrary.wiley.com/doi/abs/10.1111/0022-1082.00101},
  eprint = {https://onlinelibrary.wiley.com/doi/pdf/10.1111/0022-1082.00101},
  year = {1999}
}

@article{10.1093/rfs/hhab055,
  author = {Medhat, Mamdouh and Schmeling, Maik},
  title = {Short-term Momentum},
  journal = {The Review of Financial Studies},
  volume = {35},
  number = {3},
  pages = {1480-1526},
  year = {2021},
  month = {06},
  issn = {0893-9454},
  doi = {10.1093/rfs/hhab055},
  url = {https://doi.org/10.1093/rfs/hhab055},
  eprint = {https://academic.oup.com/rfs/article-pdf/35/3/1480/42571397/hhab055.pdf},
}

@article{merton1987simple,
  title={A simple model of capital market equilibrium with incomplete information},
  author={Merton, Robert C and others},
  year={1987},
  publisher={Sloan School of Management, Massachusetts Institute of Technology;}
}

@article{Namouri26012018,
author = {Hela Namouri and Fredj Jawadi and Zied Ftiti and Néjib Hachicha},
title = {Threshold effect in the relationship between investor sentiment and stock market returns: a PSTR specification},
journal = {Applied Economics},
volume = {50},
number = {5},
pages = {559--573},
year = {2018},
publisher = {Routledge},
doi = {10.1080/00036846.2017.1335387},
URL = {https://doi.org/10.1080/00036846.2017.1335387},
eprint = {https://doi.org/10.1080/00036846.2017.1335387}
}

@article{ORTMANN2020101717,
title = {COVID-19 and investor behavior},
journal = {Finance Research Letters},
volume = {37},
pages = {101717},
year = {2020},
issn = {1544-6123},
doi = {https://doi.org/10.1016/j.frl.2020.101717},
url = {https://www.sciencedirect.com/science/article/pii/S1544612320307959},
author = {Regina Ortmann and Matthias Pelster and Sascha Tobias Wengerek}
}

@article{Ozik_Sadka_Shen_2021, 
title={Flattening the Illiquidity Curve: Retail Trading During the COVID-19 Lockdown}, 
volume={56}, 
DOI={10.1017/S0022109021000387}, 
number={7}, 
journal={Journal of Financial and Quantitative Analysis}, 
author={Ozik, Gideon and Sadka, Ronnie and Shen, Siyi}, 
year={2021}, 
pages={2356–2388}
}

@article{PENG2006563,
title = {Investor attention, overconfidence and category learning},
journal = {Journal of Financial Economics},
volume = {80},
number = {3},
pages = {563-602},
year = {2006},
issn = {0304-405X},
doi = {https://doi.org/10.1016/j.jfineco.2005.05.003},
url = {https://www.sciencedirect.com/science/article/pii/S0304405X05002138},
author = {Lin Peng and Wei Xiong}
}

@article{10.1093/rfs/hhn053,
    author = {Petersen, Mitchell A.},
    title = {Estimating Standard Errors in Finance Panel Data Sets: Comparing Approaches},
    journal = {The Review of Financial Studies},
    volume = {22},
    number = {1},
    pages = {435-480},
    year = {2008},
    month = {06},
    issn = {0893-9454},
    doi = {10.1093/rfs/hhn053},
    url = {https://doi.org/10.1093/rfs/hhn053},
    eprint = {https://academic.oup.com/rfs/article-pdf/22/1/435/24453539/hhn053.pdf},
}

@article{doi:10.1086/378531,
author = {Scheinkman, Jos\'{e} A. and Xiong, Wei},
title = {Overconfidence and Speculative Bubbles},
journal = {Journal of Political Economy},
volume = {111},
number = {6},
pages = {1183-1220},
year = {2003},
doi = {10.1086/378531},
URL = {https://doi.org/10.1086/378531},
eprint = {https://doi.org/10.1086/378531}
}

@article{10.1093/rfs/hhj032,
    author = {Statman, Meir and Thorley, Steven and Vorkink, Keith},
    title = {Investor Overconfidence and Trading Volume},
    journal = {The Review of Financial Studies},
    volume = {19},
    number = {4},
    pages = {1531-1565},
    year = {2006},
    month = {03},
    issn = {0893-9454},
    doi = {10.1093/rfs/hhj032},
    url = {https://doi.org/10.1093/rfs/hhj032},
    eprint = {https://academic.oup.com/rfs/article-pdf/19/4/1531/24421788/hhj032.pdf},
}

@article{STOFFMAN201450,
title = {Who trades with whom? Individuals, institutions, and returns},
journal = {Journal of Financial Markets},
volume = {21},
pages = {50-75},
year = {2014},
issn = {1386-4181},
doi = {https://doi.org/10.1016/j.finmar.2014.08.002},
url = {https://www.sciencedirect.com/science/article/pii/S1386418114000652},
author = {Noah Stoffman}
}

\end{document}